\documentclass[1p, authoryear]{elsarticle}

\usepackage{amsmath}
\usepackage{amssymb}
\usepackage{bbm}
\usepackage{hyperref} 

\newtheorem{theorem}{Theorem}[section]

\newtheorem{proposition}[theorem]{Proposition}

\def\bs{\boldsymbol}

\begin{document}

 \title{Lasso ANOVA Decompositions for Matrix and Tensor Data\tnoteref{t1}} \tnotetext[t1]{Supplementary material, including proofs of all propositions and additional numerical results, can be found in the online version of the paper.
A stand-alone package for implementing LANOVA penalization for matrices and three-way tensors can be downloaded from \href{https://github.com/maryclare/LANOVA}{\texttt{https://github.com/maryclare/LANOVA}}.}

\author[1]{Maryclare Griffin\corref{cor1}% 
}
 \ead{maryclare@cornell.edu}
\author[2]{Peter Hoff
} \ead{peter.hoff@duke.edu}
\cortext[cor1]{Corresponding author}

\address[1]{Center for Applied Mathematics, Cornell University, Ithaca, NY, USA }
\address[2]{Department of Statistical Science, Duke University, Durham, NC, USA}

% \author{Maryclare Griffin$^{*}$\\
%    Center for Applied Mathematics, Cornell University,\\ Ithaca, NY 14853, USA \\
%    and \\
%    Peter D. Hoff \\
%    Department of Statistical Science, Duke University, \\ Durham, NC 27710, USA
%    \blfootnote{
%    Corresponding author. Email address: \href{mailto:maryclare@cornell.edu}{maryclare@cornell.edu}}
%}

\begin{abstract} % Must be less than 225 words
Consider the problem of estimating the entries of an unknown mean matrix or tensor given a single noisy realization. In the matrix case, this problem can be addressed by decomposing the mean matrix into a component that is additive in the rows and columns, i.e.\ the additive ANOVA decomposition of the mean matrix, plus a matrix of elementwise effects, and assuming that the elementwise effects may be sparse. Accordingly, the mean matrix can be estimated by solving a penalized regression problem, applying a lasso penalty to the elementwise effects. 
Although solving this penalized regression problem is straightforward, specifying appropriate values of the penalty parameters is not. Leveraging the posterior mode interpretation of the penalized regression problem, moment-based empirical Bayes estimators of the penalty parameters can be defined.
Estimation of the mean matrix using these these moment-based empirical Bayes estimators can be called LANOVA penalization, and the corresponding estimate of the mean matrix can be called the LANOVA estimate.
The empirical Bayes estimators are shown to be consistent. Additionally, LANOVA penalization is extended to accommodate sparsity of row and column effects and to estimate an unknown mean tensor. 
The behavior of the LANOVA estimate is examined under misspecification of the distribution of the elementwise effects, and LANOVA penalization is  applied to several datasets, including a matrix of microarray data, a three-way tensor of fMRI data and a three-way tensor of wheat infection data.
\end{abstract}


  \maketitle

\section{Introduction}\label{sec:intro}

Researchers are often interested in estimating the entries of an unknown $n\times p$ mean matrix $\bs M$ given a single noisy realization, $\bs Y = \bs M + \bs Z$, where the entries of $\bs Z$ are assumed to be independent, identically distributed mean zero normal random variables with unknown variance $\sigma^2_z$.
Consider a noisy matrix $\bs Y$ of gene expression measurements for different genes and tumors. Researchers may be interested in which tumors have unique gene expression profiles, and which genes are differentially expressed across different tumors.
%Alternatively, consider a noisy three-way array $\bs Y$ of experimental data on infection of different wheat varieties with different strains of blight across several years from a three-way full factorial design. Researchers may be interested in identifying a small number of year-by-strain-by-variety interactions.

This is challenging because no replicates are observed. Each unknown $m_{ij}$ corresponds to a single observation $y_{ij}$, 
and so the maximum likelihood estimate $\bs Y$ has high variability. 
Accordingly, simplifying assumptions that reduce the dimensionality of $\bs M$ are often made.
Many such assumptions relate to a two-way ANOVA decomposition of $\bs M$:
\begin{align}\label{eq:decomp}
\bs M = \mu \bs 1_n \bs 1_p' + \bs a \bs 1_p' + \bs 1_n \bs b' + \bs C,
\end{align}
where $\mu$ is an unknown grand mean, $\bs a$ is an $n \times 1$ vector of unknown row effects, $\bs b$ is a $p\times 1$ vector of unknown column effects, $\bs C$ is a matrix of elementwise ``interaction'' effects 
and $\bs 1_n$ and $\bs 1_p$ are $n\times 1$ and $p \times 1$ vectors of ones, respectively.
In the absence of replicates, implicitly assuming $\bs C = \bs 0$ is common.
This reduces the number of freely varying unknown parameters, from $np$ to $n + p$, but is also unlikely to be appropriate in practice.

Alternatively, one might assume that elements of $\bs C$ can be written as a function of a small number $R$ of multiplicative components, i.e. $c_{ij} = \sum_{r= 1}^R u_{r,i} v_{r,j}$ where $\bs u_r$ and $\bs v_r$ and $n \times 1$ and $p\times 1$ row and column factors. This corresponds to a low-rank matrix $\bs C$ and an additive-plus-low-rank mean matrix $\bs M$. Additive-plus-low-rank models have a long history and continue to be very popular \citep{Fisher1923,Gollob1968, Johnson1972b, Mandel1971, Goodman1990, Forkman2014}. 
However, in the settings we consider it is reasonable to expect that $\bs C$ may be sparse with a relatively small number of nonzero elements, e.g.\ some tumor-gene combinations may have large interaction effects while others may have negligible interaction effects. In such settings, it is easy to imagine scenarios in which a low-rank estimate of $\bs M$ may fail, e.g.\ if $\bs Y$ were an $n\times n$ square matrix and all $c_{ii}$ were large while all $c_{ij}$, $i\neq j$ were equal to zero. In this case, a low-rank estimate of $\bs C$ would not suffice because a full rank $R = n$ estimate  would be needed.
%Assuming that $\bs C$ is rank $R$ implies that elements of.
%
%Although useful in many settings, assuming low rank $\bs C$ has the two main limitations. 
%First, in the presence of unknown noise variance, $\sigma^2_z$, existing methods for choosing the rank can be computationally expensive for large matrices \citep{Hoff2007}, require strong assumptions such as known $\sigma^2_z$ \citep{Candes2013},  or rely on approximations to account for unknown $\sigma^2_z$ that may not always perform well in practice \citep{Josse2016}. 
%%XXX \cite{Forkman2014} provides a recent review of difficulties that arise when choosing the rank of $\bs C$.
%Second, even when the rank can be chosen well, these methods conflate the presence of elementwise effects of scientific interest with the presence of multiplicative effects.
%While this may be plausible in many settings, 

If $\bs M$ is approximately additive in the sense that large deviations from additivity are rare, then $\bs C$ is sparse and estimation of $\bs M$ may be improved by penalizing elements of $\bs C$:
\begin{align}\label{eq:obj}
\text{min}_{\mu,\bs a,\bs b, \bs C}\frac{1}{2\sigma^2_z}\left|\left|\text{vec}\left\{\bs Y - \left( \mu \bs 1_n \bs 1_p' + \bs a \bs 1_p' + \bs 1_n \bs b' + \bs C\right)\right\} \right|\right|^2_2 + \lambda_c \left|\left| \text{vec}\left(\bs C\right)\right|\right|_1. 
\end{align}
The $\ell_1$ penalty induces sparsity among the estimated entries of $\bs C$ and solving this penalized regression problem yields unique estimates of $\bs M$ and $\bs C$. 
Elements of $\bs C$ can be interpreted as interactions insofar as they indicate deviation from a strictly additive model.

Although  \eqref{eq:obj} is a standard lasso regression problem that can be solved easily given values of $\lambda_c$ and $\sigma^2_z$, specifying values of $\lambda_c$ and $\sigma^2_z$ is uniquely challenging in this setting. 
The methods suggested by \cite{Donoho1994} and \cite{Donoho1995}  are not appropriate because they are specific to orthogonal design matrices; columns of the design matrix corresponding to the regression problem in \eqref{eq:obj} are correlated. The same is true of the unbiased risk estimate minimization procedure suggested by \cite{Tibshirani1996}.
Although columns of the design matrix will become less correlated as $n$ and $p\rightarrow \infty$, the correlations may not be negligible in practice especially if $n$ or $p$ is relatively small.
\cite{Tibshirani1996} also suggested cross validation, which could be performed after rewriting Equation~\eqref{eq:obj} to depend on a single parameter $\eta = \lambda_c \sigma^2_z$. 
However, cross-validation is also poorly suited to this setting. Consider leave-one-out cross validation to select a value of $\eta$, and suppose we hold out $y_{11}$ and solve \eqref{eq:obj} for any fixed value of $\eta$ using the elements of $\bs Y$ excluding $y_{11}$. We obtain estimates of $\mu$, $\bs a$, $\bs b$, and all elements of $\bs C$ \emph{except} $c_{11}$, as only the held out test data point $y_{11}$ contains any information about $c_{11}$. For this reason, we cannot compute an out-of-sample prediction for $y_{11}$ without making additional assumptions that relate $\mu$, $\bs a$, $\bs b$ and all of the elements of $\bs C$ except $c_{11}$ to $c_{11}$ and selecting $\eta$ by cross validation without additional assumptions is not possible.

This penalized regression problems has been considered in the literature on outlier detection, as nonzero elements of $\bs C$ can alternatively be interpreted as outliers. 
\cite{She2011} interpret elements of $\bs C$ in this way and consider the more general problem with an arbitrary full rank design matrix $\bs X$.
They approach specification of $\lambda_c$ and $\sigma^2_z$ by introducing a conservative extension of the methods suggested by \cite{Donoho1994} for orthogonal design matrices, setting $np$ different values  $\lambda_i = \sigma\sqrt{2\left(1 - h_{ii}\right)\text{log}\left(np\right)}$ where $\bs H = \bs X \left(\bs X'\bs X\right)^{-1}\bs X$. Because $\sigma^2$ can be very challenging to estimate, they suggest setting $\lambda_i = \lambda \sqrt{1 - h_{ii}}$ in a data-adaptive way using a modified BIC.
Although the methods proposed by \cite{She2011} have the advantage of applying to general regression problems with arbitrary design matrices $\bs X$, they have computational disadvantages in high dimensions because they require computing an initial robust estimate of and iteratively re-estimating $\bs C$.

%The crux of the problem is that there is one unknown mean parameter $c_{ij}$ per observation $y_{ij}$, and \emph{only} observation $y_{ij}$ contains information about $c_{ij}$.
We take another approach and view the $\ell_1$ penalty on $\bs C$ as a Laplace 
prior distribution, in which case $\lambda_c$ and $\sigma^2_z$ can be interpreted as nuisance parameters that can be estimated from the data.
The relationship between the $\ell_1$ penalty and the Laplace prior has long been acknowledged \citep{Tibshirani1996}. It offers not only a framework for specifying $\lambda_c$ and $\sigma^2_z$, but also decision theoretic justifications for using estimates of $\bs M$ and $\bs C$ obtained by solving \eqref{eq:obj} using estimated $\lambda_c$ and $\sigma^2_z$ because the posterior mode is known to minimize a specific data-adaptive loss function \citep{Pratt1965, Tiao1973}.
The challenge is in the estimation of $\lambda_c$ and $\sigma^2_z$, because  computing maximum marginal likelihood estimates may be prohibitively computationally demanding and intractable in practice
\citep{Figueiredo2003, Park2008}. 
In this paper we instead present moment-based empirical Bayes estimators of the nuisance parameters $\lambda_c$ and $\sigma^2_z$ that 
are easy to compute, consistent and 
independent of assumptions made regarding $\bs a$ and $\bs b$. 
%Moment-based estimators can be sensitive to outliers, however we consider high dimensional settings in which moments can be estimated well.
As our approach to estimating $\lambda_c$ and $\sigma^2_z$ uses the Laplace prior interpretation of the $\ell_1$ penalty, we refer to estimation of $\bs M$ via optimization of 
Equation~\eqref{eq:obj} using these nuisance parameter estimators as LANOVA penalization and we refer to the estimate $\widehat{\bs M}$ as the LANOVA estimate.

%LANOVA penalization in particular should only be used when the context of the specific problem suggests that the noise is normally distributed, e.g. when elements of $\bs Y$ are sample means.

The paper proceeds as follows:
In Section~\ref{sec:lapen}, we introduce moment-based  estimators for $\lambda_c$ and $\sigma^2_z$, show that they are consistent as \emph{either} the number of rows \emph{or} columns of $\bs Y$ go to infinity. We show that their efficiency is comparable to that of asymptotically efficient marginal maximum likelihood estimators (MMLEs). 
In Section~\ref{sec:laimp}, we discuss estimation of $\bs M$ via Equation~\eqref{eq:obj} given estimates of $\lambda_c$ and $\sigma^2_z$ and introduce a test of whether or not elements of $\bs C$ are heavy-tailed, which allows us to avoid LANOVA penalization in settings where it is especially inappropriate.
%Specifically, we discuss how population average interaction effects can be obtained from the LANOVA estimate of $\bs M$.
%Additionally, we introduce a test of whether or not elements of $\bs C$ are heavy-tailed, which is essentially a test of whether $\bs C$ and $\bs Z$ can be ``deconvoluted.'' Although this is not a test of the Laplace assumption for elements of $\bs C$ specifically, it allows us to avoid LANOVA penalization in settings where it is especially inappropriate.
We also investigate the performance of LANOVA estimates of $\bs M$ relative to strictly additive estimates, strictly non-additive estimates, additive-plus-low-rank estimates, IPOD estimates from \cite{She2011}, and approximately minimax estimates based on \cite{Donoho1994}, and examine robustness to misspecification of the distribution of elements of $\bs C$.
In Section~\ref{sec:ext}, we extend LANOVA penalization to include penalization of lower-order mean parameters $\bs a$ and $\bs b$ and also to apply to the case where $\bs Y$ and $\bs M$ are $K$-way tensors. 
In Section~\ref{sec:numex}, we apply LANOVA penalization to a matrix of gene expression measurements, a three-way tensor of fMRI data and a three-way tensor of wheat infection data. 
%For the matrix of microrray data, we use LANOVA penalization to identify genes that are differentially expressed across different brain tumors and brain tumors that display unique gene expression profiles.
%For the tensor of fMRI data, we use LANOVA penalization to perform an exploratory analysis of spatial patterns in activation response to different tasks over time.
%For the tensor of wheat infection data, we use LANOVA penalization to identify ``real'' third order interaction effects from a full factorial experiment.
In Section~\ref{sec:disc} we discuss extensions, specifically multilinear regression models and opportunities that arise in the presence of replicates.

\section{LANOVA Nuisance Parameter Estimation}\label{sec:lapen}

Consider the following statistical model for deviations % $\bs C+\bs Z$
of $\bs Y$ from a strictly additive model:
\begin{align}\label{eq:matmod}
\bs Y &  = \mu \bs 1_n \bs 1_p' + \bs a \bs 1_p' + \bs 1_n \bs b' + \bs C + \bs Z,  \\
\bs C & = \{ c_{ij} \}  \sim  \text{i.i.d.\ Laplace}\left(0,\lambda_c^{-1} \right)\text{, \quad} \bs Z = \{ z_{ij} \}  \sim  \text{i.i.d.\ }N\left(0,\sigma^2_z \right).  \nonumber
\end{align}
The posterior mode of $\mu$, $\bs a$, $\bs b$ and $\bs C$ under this Laplace prior for $\bs C$ 
and flat priors for $\mu$, $\bs a$ and $\bs b$
corresponds to the solution of LANOVA penalization problem 
given by Equation~\eqref{eq:obj}. 
%The problem of specifying the tuning parameters for LANOVA penalization is equivalent to estimation of the unknown parameters $\lambda_c$ and $\sigma^2_z$.
%The posterior mode
%consists of an additive part $\bs A$
%and a  possibly sparse matrix $\bs C$ of deviations from additivity. 
%We will obtain adaptive posterior mode estimates whereby values of $\sigma^2_z$ and $\lambda_c$ 
%are estimated from the data and then used to specify the objective function \eqref{eq:obj}. 
%Doing so requires estimates of $\lambda_c$ and $\sigma^2_z$.
%Rather than construct our estimates of $\lambda_c$ and $\sigma^2_z$ directly from $\bs Y$, we suppose for a moment that no prior distribution were assumed for elements of $\bs C$.
%In this case, an OLS estimate of $\bs C + \bs Z$ would be given by $\bs R = \bs H_n \bs Y \bs H_p$, where $\bs H_n = \bs I_n - \bs 1_n \bs 1_n/n$ is an orthogonal projection matrix, i.e. $\bs H_n = \bs H_n'$, $\bs H_n \bs H_n = \bs H_n$ and $\bs 1_n \bs H_n = \bs H_n \bs Y \bs H_p$, and $\bs H_p$ is defined accordingly. 
%The matrix $\bs R$ is a very useful quantity. It not only contains information about both $\lambda_c$ and $\sigma^2_z$, but also is independent of $\mu$, $\bs a$ and $\bs b$: 

We construct estimators of $\lambda_c$ and $\sigma^2_z$ as follows.
Letting $\bs H_k = \bs I_k - \bs 1_k \bs 1_k/k$ be the $k\times k$ 
centering matrix, we define $\bs R = \bs H_n \bs Y \bs H_p$. $\bs R$ dependes on $\bs C$ and $\bs Z$ alone, specifically $\bs R =  \bs H_n \left(\bs C + \bs Z\right) \bs H_p$.
We construct estimators of $\lambda_c$ and $\sigma^2_z$ from $\bs R$ by leveraging the difference between Laplace and normal tail behavior as measured by fourth order moments. 
The fourth order central moment of any random variable $x$ with mean $\mu_x$ and variance $\sigma^2_x$ can be 
expressed as  $\mathbb{E}\left[\left(x - \mu_x\right)^4 \right] = \left(\kappa + 3\right)\sigma^4_x$, where $\kappa$ is interpreted as the \emph{excess kurtosis} of the distribution of $x$ relative to a normal distribution.
A normally distributed variable has excess kurtosis equal to $0$, whereas a Laplace distributed random variable has excess kurtosis equal to $3$. It follows that the second and fourth order central moments of elements of $\bs C + \bs Z$ are $\mathbb{E}\left[\left(c_{ij} + z_{ij}\right)^2 \right] = \sigma^2_c + \sigma^2_z$ and $\mathbb{E}\left[\left(c_{ij} + z_{ij}\right)^4 \right] = 3\sigma^4_c + 3\left(\sigma^2_c + \sigma^2_z \right)^2$, respectively, 
where $\sigma^2_c = 2/\lambda^2_c$ is the variance  of a Laplace$(0, \lambda_c^{-1})$ random variable. 
Given values of $\mathbb{E}\left[\left(c_{ij} + z_{ij}\right)^2\right]$ and $\mathbb{E}\left[\left(c_{ij} + z_{ij}\right)^4\right]$, we see that $\sigma^2_c$ and $\sigma^2_z$, and accordingly $\lambda_c$, can easily be recovered.

We do not observe $\bs C + \bs Z$ directly, but we can use the the second and fourth order sample moments of $\bs R$, 
an estimate of $\bs C + \bs Z$, given by $\overline{r}^{\left(2\right)} = \frac{1}{np}\sum_{i = 1}^n \sum_{j = 1}^p r^2_{ij}$ and $\overline{r}^{\left(4\right)} = \frac{1}{np}\sum_{i = 1}^n \sum_{j = 1}^p r^4_{ij}$, respectively, to separately estimate $\sigma^2_c$ and $\sigma^2_z$. These estimators are:
\begin{align} \label{eqn:momest}
\widehat{\sigma}^4_c =& \left\{\frac{n^3p^3}{\left(n - 1\right)\left(n^2 - 3n + 3\right)\left(p - 1\right)\left(p^2 - 3p + 3\right)}\right\}\left\{\overline{r}^{\left(4\right)}/3 - \left(\overline{r}^{\left(2\right)}\right)^2 \right\}, \\ \nonumber 
\widehat{\sigma}^2_c =& \sqrt{\widehat{\sigma}^4_c}\text{,\quad} \widehat{\sigma}^2_z = \left\{\frac{np}{\left(n - 1\right)\left(p - 1\right)}\right\}\overline{r}^{\left(2\right)} - \widehat{\sigma}^2_c.
\end{align}
An estimator of $\lambda_c$ is then given by $\widehat{\lambda}_c = \sqrt{2/\widehat{\sigma}^2_c}$. Studying the properties of these estimators is slightly challenging, as elements of $\bs R$ are neither independent nor identically distributed.
 
%It follows that $\widehat{\lambda}_c = \sqrt{\frac{2}{\widehat{\sigma}^2_c}}$ is an estimate of $\lambda_c$.

The estimator $\widehat{\sigma}^4_c$ is biased. It is possible to obtain an unbiased estimator for $\sigma^4_c$, however the unbiased estimator will not be consistent as $n\rightarrow \infty$ with $p$ fixed or $p \rightarrow \infty$ with $n$ fixed. 
Because these estimators depend on higher-order terms which can be very sensitive to outliers, it is desirable to have consistency as either the number of rows or  columns grows.
Accordingly, we prefer the biased estimator and examine its bias in the following proposition.
\begin{proposition}\label{prop:bias}
Under the model given by Equation~\eqref{eq:matmod}, 
\begin{align*}
\mathbb{E}\left[\widehat{\sigma}^4_c\right] -\sigma^4_c =& -\left\{\frac{n^3 p^3}{\left(n - 1\right)\left(n^2 - 3n + 3\right)\left(p - 1\right)\left(p^2 - 3p + 3\right)}\right\}\left[\left\{\frac{3\left(n - 1\right)^2\left(p - 1\right)^2}{n^3p^3}\right\}\sigma^4_c +\right.\\
&\left. \left\{\frac{2\left(n - 1\right)\left(p - 1\right)}{n^2p^2}\right\}\left(\sigma^2_c + \sigma^2_z\right)^2 \right].
\end{align*}
\end{proposition} % comment on the possibility or difficulty of unbiased estimation
A proof of this proposition and all other results presented in this paper are given in an web appendix. 
The bias is always negative and accordingly, yields  overpenalization of $\bs C$. 
When both $n$ and $p$ are small, $\widehat{\sigma}^4_c$ tends to underestimate $\sigma^4_c$.
Recalling that $\sigma^4_c$ is inversely related to $\lambda_c$, this reflects a tendency to overpenalize and accordingly overshrink elements of $\bs C$ when both $n$ and $p$ are small.
This is desirable, in that it reflects a tendency to prefer the simple additive model when few data are available. % Discuss 
We also observe that the bias depends on both $\sigma^2_c$ and $\sigma^2_z$.
Holding $n$, $p$ and $\sigma^2_c$ fixed, we will overestimate $\lambda_c$ more when $\sigma^2_z$ is larger.
Again, this is desirable, in that it reflects a tendency to prefer the simple additive model when the data are very noisy.
Last, we see that the bias is $O\left(1/np\right)$, i.e.\ the bias approaches zero as \emph{either} the number of rows \emph{or} the number of columns increases. The large sample behavior of our estimators of the nuisance parameters is similar.
\begin{proposition}\label{prop:consestsmat}
Under the model given by Equation~\eqref{eq:matmod}, $\widehat{\sigma}^4_c\stackrel{p}{\rightarrow}\sigma^4_c$, $\widehat{\sigma}^2_c\stackrel{p}{\rightarrow}\sigma^2_c$, $\widehat{\lambda}_c\stackrel{p}{\rightarrow}\lambda_c$ and $\widehat{\sigma}^2_z\stackrel{p}{\rightarrow}\sigma^2_z$ as $n \rightarrow \infty$ with $p$ fixed, $p\rightarrow \infty$ with $n$ fixed, or $n, p\rightarrow \infty$.
\end{proposition}

\begin{figure}[h]
\centering
\includegraphics[scale = 0.55]{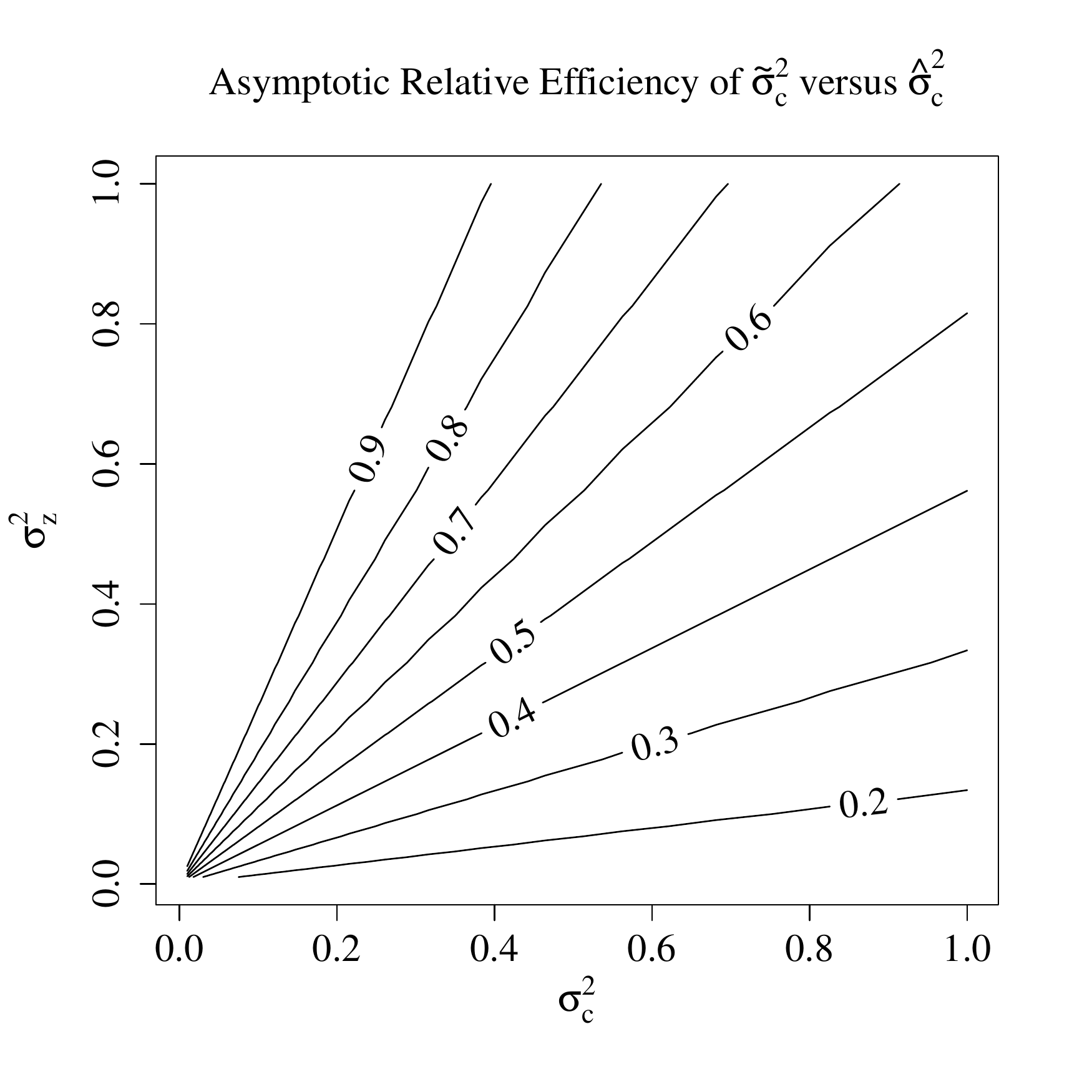}
\caption{Asymptotic relative efficiency $\mathbb{V}[\widetilde{\sigma}^2_c]/\mathbb{V}[\widehat{\sigma}^2_c]$ of the MMLE $\widetilde{\sigma}^2_c$ versus our moment-based estimator $\widehat{\sigma}^2_c$ as a function of the true variances $\sigma^2_c$ and $\sigma^2_z$.} \label{fig:comptomle}
\end{figure}

Although these nuisance parameter estimators are easy to compute and consistent as $n$ or $p\rightarrow \infty$, they are not maximum likelihood estimators and  may not be asymptotically efficient even as $n$ and $p\rightarrow \infty$.
Accordingly, we compare the asymptotic efficiency of our estimator $\widehat{\sigma}^2_c$ to that of the corresponding asymptotically efficient marginal maximum likelihood estimator (MMLE) denoted by $\widetilde{\sigma}^2_c$ as $n$ and $p\rightarrow \infty$. As noted in the Introduction, obtaining $\widetilde{\sigma}^2_c$ is computationally demanding because maximizing the marginal likelihood of the data requires a Gibbs-within-EM algorithm that can be slow to converge \citep{Park2008}.  
Fortunately, computing the asymptotic variance of $\widetilde{\sigma}^2_c$ is simpler than computing $\widetilde{\sigma}^2_c$ itself. 
The asymptotic variance of $\widetilde{\sigma}^2_c$ is given by the Cram\'er-Rao lower bound for $\sigma^2_c$, which can be computed numerically from the density of the sum of Laplace and normally distributed variables  \citep{Nadarajah2006,Diaz-Frances2008}.
The asymptotic variance of $\widehat{\sigma}^2_c$ is straightforward to compute as $\sqrt{np}\left(\widehat{\sigma}^4_c - \sigma^4_c\right)$ converges in distribution to a moment estimator of $\sigma^4_c$. 
We note that the asymptotic variance of $\widehat{\lambda}_c$ is similarly straightforward to compute; both asymptotic variances are given in the web appendix.
 % This may be even smaller than it would be in practice because it corresponds to the submodel where a = 0, b =0, c = 0.  Worth mentioning?
Letting $\mathbb{V}\left[\widetilde{\sigma}^{2}_c\right]$ and $\mathbb{V}\left[\widehat{\sigma}^{2}_c\right]$ refer to the variances of the estimators $\widetilde{\sigma}^{2}_c$ and $\widehat{\sigma}^{2}_c$, we plot the asymptotic relative efficiency $\mathbb{V}\left[\widetilde{\sigma}^{2}_c\right]/\mathbb{V}\left[\widehat{\sigma}^2_c\right]$ over values of $\sigma^2_c, \sigma^2_z \in \left[0, 1\right]$ in Figure~\ref{fig:comptomle}. Note that the relative efficiency of  $\widehat{\sigma}^2_c$ compared to $\widetilde{\sigma}^2_c$ also reflects the relative efficiency of our estimators $\widehat{\lambda}_c$ and $\widehat{\sigma}^2_z$ compared to the MMLEs $\widetilde{\lambda}_c$ and $\widetilde{\sigma}^2_c$, respectively, because both are simple functions of $\widehat{\sigma}^2_c$.

When $\sigma^2_c$ is small relative to $\sigma^2_z$, the MMLE $\widetilde{\sigma}^{2}_c$ tends to be slightly more efficient.
When $\sigma^2_c$ is large relative to $\sigma^2_z$, $\widetilde{\sigma}^{2}_c$ tends to be much more efficient. However, in such cases the interactions will not be heavily penalized and LANOVA penalization will not tend to yield a simplified, nearly additive estimate of $\bs M$.
Put another way, Figure~\ref{fig:comptomle} indicates that $\widehat{\lambda}_c$ and $\widehat{\sigma}^2_z$ will be nearly as efficient as the corresponding MMLEs when LANOVA penalization is useful for producing a simplified, nearly additive estimate of $\bs M$ with sparse interactions.
We also note that because they are moment-based, our estimators may be more robust to misspecification of the distribution of elements of $\bs C$ and $\bs Z$ than the MMLEs.

\section{Mean Estimation, Interpretation, Model Checking and Robustness}\label{sec:laimp}

\subsection{Mean Estimation}

In practice, our nuisance parameter estimators are not guaranteed to be nonnegative and two special cases can arise.
When $\widehat{\sigma}^4_c < 0$, we set $\widehat{\sigma}^2_c = 0$, $\widehat{\bs C} = \bs 0$, and $\widehat{\bs M} = \widehat{\bs M}_{ADD}$, where $\widehat{\bs M}_{ADD} = \left(\bs I_n - \bs H_n \right)\bs Y \left(\bs I_p - \bs H_p\right)$ is the strictly additive estimate.
When $\widehat{\sigma}^2_z < 0$, we reset $\widehat{\sigma}^2_z = 0$ and set $\widehat{\bs M} = \widehat{\bs M}_{MLE}$, where $\widehat{\bs M}_{MLE} = \bs Y$ is the strictly non-additive estimate.
Neither special case prohibits estimation of $\bs M$. 

We assess how often these special cases arise via a small simulation study. Setting $\sigma^2_z = 1$, $n = p = 25$, $\mu = 0$, $\bs a = \bs 0$ and $\bs b = \bs 0$, we simulate $10,000$ realizations of $\bs Y = \bs C + \bs Z$ under the model given by Equation~\eqref{eq:matmod} for each value of $\sigma^2_c \in \left\{1/2, 1, 3/2\right\}$. We obtain $\widehat{\sigma}^2_c \leq 0$  in 13.7\%, 1.64\% and 0.02\% of simulations for  $\sigma^2_c$ equal to $1/2$, $1$ and $3/2$, respectively. This means that when the magnitude of elements of $\bs C$ is smaller, we are more likely to obtain a strictly additive estimate of $\bs M$.
%Unsurprisingly, we are more likely to obtain $\widehat{\sigma}^2_c \leq 0$ when the data generating value of $\sigma^2_c$ is smaller relative to $\sigma^2_z$. 
We do not obtain $\widehat{\sigma}^2_z = 0$ in any simulations.

% Make this a new paragraph and then add the block coordinate descent algorithm
When $\widehat{\sigma}^2_c > 0$ and $\widehat{\sigma}^2_z  > 0$, we can obtain an estimate of $\bs M$ from Equation~\eqref{eq:obj} using block coordinate descent. Setting $\widehat{\bs C}^{0} = \bs H_n \bs Y \bs H_p$ and $k = 1$, our block coordinate descent algorithm iterates the following until the objective function Equation~\eqref{eq:obj} converges:
\begin{itemize}
	\item Set $\widehat{\mu}^{k} = \bs 1_n'(\bs Y - \widehat{\bs C}^{k-1})\bs 1_p/np$, $\widehat{\bs a}^{k} = \bs H_n'(\bs Y - \widehat{\bs C}^{k-1})\bs 1_p/p$, \\
	$\widehat{\bs b}^{k} = \bs H_p(\bs Y - \widehat{\bs C}^{k-1})'\bs 1_n/n$ and $\bs R^{k} = \bs Y - \widehat{\mu}^{k} \bs 1_n \bs 1_p' - \widehat{\bs a}^{k}\bs 1_p' - \bs 1_n (\widehat{\bs b}^{k})'$;
	\item Set $\widehat{\bs C}^{k} = \text{sign}(\bs R^{k} ) (|\bs R^{k}| - \widehat{\lambda}_c \widehat{\sigma}^2_z )_+$, where $\widehat{\lambda}_c = \sqrt{2/\widehat{\sigma}^2_c}$ $\text{sign}(\cdot)$ and the soft-thresholding function $(\cdot )_+$ are applied elementwise. Set $k = k + 1$.
\end{itemize}

\subsection{Interpretation}

The nonzero entries of $\widehat{\bs C}$ correspond to the $r$ largest residuals from fitting a strictly additive model with $\bs C = \bs 0$, where $r$ is determined by $\widehat{\lambda}_c$ and $\widehat{\sigma}^2_z$. 
Elements of $\widehat{\bs C}$ can be interpreted as interactions insofar as they indicate deviation from a strictly additive model for $\bs M$.
However, because we do impose the standard ANOVA zero-sum constraints, we cannot interpret elements of $\widehat{\bs C}$ directly as population average interaction effects, i.e.\ $\widehat{c}_{ij} \neq \mathbb{E}\left[y_{ij}\right] -\frac{1}{p} \sum_{j = 1}^p \mathbb{E}\left[y_{ij}\right] - \frac{1}{n}\sum_{i = 1}^n \mathbb{E}\left[y_{ij}\right] + \frac{1}{np}\sum_{i = 1}^n \sum_{j = 1}^p \mathbb{E}\left[y_{ij}\right]$.  For the same reason, $\mu$, $\bs a$ and $\bs b$ cannot be interpreted as the grand mean and population average main effects.
To obtain estimates that have the standard population average interpretation, we recommend performing a two-way ANOVA decomposition of $\widehat{\bs M}$. 
In the appendix, we show that the grand mean and population average main effects obtained via ANOVA decomposition of $\widehat{\bs M}$ are identical to those obtained by performing an ANOVA decomposition of $\bs Y$.

\subsection{Testing}

LANOVA penalization assumes the distribution of entries of $\bs C$ have tail behavior consistent with a Laplace distribution.
It is natural to ask if this assumption is appropriate, but it is difficult to test it because $\bs C$ and $\bs Z$ enter into the observed data through their sum $\bs C + \bs Z$. 
Accordingly, we suggest a test of the more general assumption that elements of $\bs C$ are heavy-tailed.
This allows us to rule out LANOVA penalization when it is especially inappropriate, i.e. when the data suggest elements of $\bs C$ are normal tailed.
When the distribution of elements of $\bs C$ is heavy-tailed, the distribution of elements of $\bs C + \bs Z$ will also be heavy-tailed and will have strictly positive excess kurtosis.
In contrast, when elements of $\bs C$ are either all zero or have a distribution with normal tails, elements of $\bs C + \bs Z$ will have excess kurtosis equal to exactly zero.
We construct a test of the  null hypothesis $H_0$: $c_{ij} + z_{ij} \sim \text{i.i.d.}\ N\left(0,\sigma^2_c + \sigma^2_z\right)$, which encompasses the cases in which $\bs C = \bs 0$ or elements of $\bs C$ are normally distributed.
Conveniently, the test statistic is a simple function of $\widehat{\sigma}^2_c$ and $\widehat{\sigma}^2_z$ and can be computed at little additional computational cost. 
We can also think of this as a test of deconvolvability of $\bs C + \bs Z$, where the null hypothesis is that deconvolution of $\bs C + \bs Z$ is not possible.

\begin{proposition}\label{prop:test}
For $\bs Y = \mu \bs 1_n \bs 1_p' + \bs a \bs 1_p' + \bs 1_n \bs b' + \bs C + \bs Z$, as $n$ and $p\rightarrow \infty$ an asymptotically level-$\alpha$ test of
$H_0$: $c_{ij} + z_{ij} \sim \text{i.i.d.}\ N\left(0,\sigma^2_c + \sigma^2_z\right)$
% the null hypothesis, elements of $\bs C + \bs Z$ are independent, identically distributed mean zero normal random variables 
%, against the alternative hypothesis, the distribution of elements of $\bs C + \bs Z$ has heavier than normal tails 
is obtained by rejecting $H_0$ when 
\begin{align*}
\sqrt{np}\left\{\frac{\widehat{\sigma}^4_c}{\sqrt{\frac{8}{3}}\left(\widehat{\sigma}^2_c + \widehat{\sigma}^2_z\right)^2}\right\} > z_{1 - \alpha},
\end{align*}
where $z_{1-\alpha}$ denotes the $1-\alpha$ quantile of the standard normal distribution.
\end{proposition}
This test gives us power against the alternative where elements $\bs C$ are heavy-tailed and LANOVA penalization may be appropriate.

Because this is an approximate test, we assess its level in finite samples in a small simulation study.
Setting $\sigma^2_z = 1$, $n = p$, $\mu = 0$, $\bs a = \bs 0$ and $\bs b = \bs 0$, we simulate $10,000$ realizations of $\bs Y = \bs C + \bs Z$ under $H_0$ for each value of $n \in \left\{25, 100\right\}$ and $\sigma^2_c \in \left\{1/2, 1, 3/2\right\}$. When $n = p = 25$, the test rejects at a slightly higher rate than the nominal level. It rejects in 7.98\%, 7.65\% and 8.66\% of simulations for $\sigma^2_c$ equal to $1/2$, $1$ and $3/2$, respectively. When $n = p = 100$, the test nearly achieves the desired level. It rejects in 6.13\%, 5.60\% and 6.00\% of simulations for $\sigma^2_c$ equal to $1/2$, $1$ and $3/2$, respectively.  We compute the approximate power of this test under two 
heavy-tailed distributions for elements of $\bs C$: the Laplace distribution assumed for LANOVA penalization and a Bernoulli-normal spike-and-slab distribution.

\begin{figure}[h]
\centering
\includegraphics[scale=0.85]{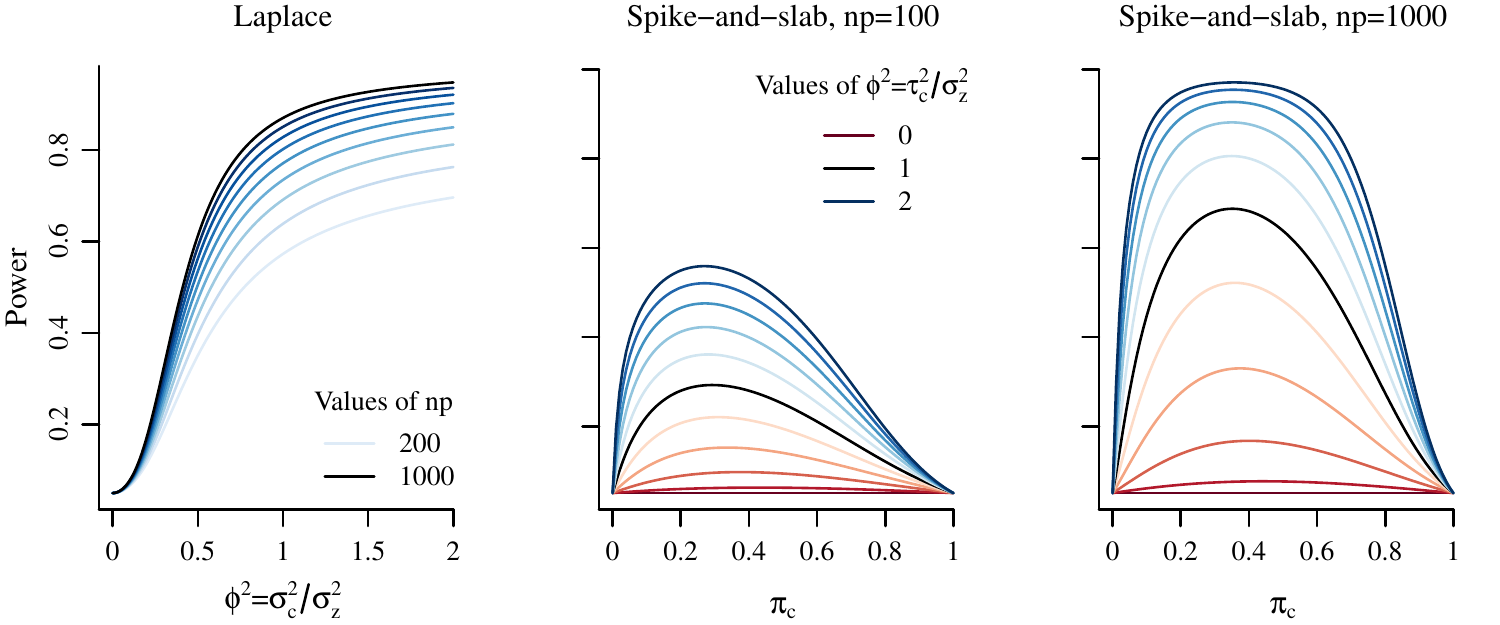}
\caption{Approximate power of the test described in Proposition~\ref{prop:test}.}
\label{fig:power}
\end{figure}

%First, we compute the approximate power of the test under a Laplace distribution for elements of $\bs C$.
\begin{proposition}\label{prop:power_lap}
Assume that elements of $\bs C$ are independent, identically distributed mean zero Laplace random variables with variance $\sigma^2_c$ and let $\phi^2 = \sigma^2_c/\sigma^2_z$. Then as $n$ and $p\rightarrow \infty$, the asymptotic power of the test given by Proposition~\ref{prop:test} is: 
\begin{align*}
1 - \Phi\left[\frac{z_{1 - \alpha} - \sqrt{\frac{3np}{8}}\left(\frac{\phi^2}{\phi^2 + 1}\right)^2}{\sqrt{1 + \left\{\frac{68\phi^8 + 36\phi^6 + 9\phi^4}{\left(1 + \phi^2\right)^4}\right\}}}\right].
\end{align*}
\end{proposition}
The power depends on the variances  $\sigma^2_c$
and $\sigma^2_z$ only through their ratio $\phi^2$.
%When $\phi^2 = 0$, i.e. when $\sigma^2_c = 0$ and all elements $\bs C$ are exactly $0$, the approximate power is equal to $\alpha$.
%This is because elements of $\bs C + \bs Z$ are normally distributed when elements of $\bs C$ are all exactly equal to zero.
It is plotted for for $\alpha = 0.05$, $\phi^2 \in \left[0,2\right]$ and $np = \left\{100, 200, \dots, 1000\right\}$ 
 in Figure~\ref{fig:power}. 
 The power of the test is increasing in $\phi^2$ and increasing more quickly when $np$ is larger and more data are available.  

Now we consider the power for Bernoulli-normal distributed elements of $\bs C$.
\begin{proposition}\label{prop:power_ss}
Assume that elements of $\bs C$ are independent, identically distributed Bernoulli-normal random variables. An element of $\bs C$ is  exactly equal to zero with probability $1 - \pi_c$, and normally distributed with mean zero and variance $\tau^2_c$ otherwise. Letting $\phi^2 = \tau^2_c/\sigma^2_z$, as $n$ and $p\rightarrow \infty$, the asymptotic power of the test given by Proposition~\ref{prop:test} is: 
\begin{align*}
1 - \Phi\left[\frac{z_{1 - \alpha} - \pi_c \left(1 - \pi_c\right)\left\{\sqrt{\frac{3np}{8}}\left(\frac{\phi^2}{\pi_c \phi^2 + 1}\right)^2\right\}}{\sqrt{1 + \pi_c \left(1 - \pi_c\right)\left\{\frac{\left(20\pi_c^2 - 28\pi_c + 35\right)\phi^8 + 16\left(5 - \pi_c\right)\phi^6 + 72\phi^4}{8\left(\pi_c \phi^2 + 1\right)^4}\right\}}}\right].
\end{align*}
\end{proposition}

The approximate power depends on the variances of the nonzero effects $\tau^2_c$ and the noise $\sigma^2_z$ only through their ratio $\phi^2$.
%We also observe that when $\pi_c = 0$, $\pi_c = 1$ or $\phi^2 = 0$, i.e. $\tau^2_c = 0$, the approximate power of the test is $\alpha$.
%Elements of $\bs C + \bs Z$ are normally distributed in these cases.
%Again, additional properties of the approximate power of the test under this alternative are difficult to interpret directly, so we plot 
It is plotted for $\alpha = 0.05$, $\pi_c \in \left[0,1\right]$, $\phi^2 \in \left\{0, 0.2, \dots, 2\right\}$ and $np = \left\{100, 1000\right\}$ in Figure~\ref{fig:power}.
The approximate power is always increasing in $\phi^2$ and $np$.
For fixed $\phi^2$ and $np$, power diminishes as the probability of an element of $\bs C$ being nonzero $\pi_c$ approaches $0$ or $1$ and $\bs C + \bs Z$ becomes more normally distributed.
%The approximate power diminishes more rapidly as $\pi_c$ approaches $1$ then as $\pi_c$ approaches $0$, especially when $\phi^2$ or $np$ is small, which is what we would expect given that the distribution of the elements of $\bs C$ becomes more normal as more elements of $\bs C$ are nonzero.
Overall, the test is more powerful when estimating $\bs C$ separately from $\bs Z$ is  more valuable, e.g.\ when elements of $\bs C$ are large in magnitude relative to the noise and when many entries of $\bs C$ are exactly zero. 

\subsection{Robustness and Comparative Performance}

%When the test rejects, it does \emph{not} necessarily suggest that the distribution of elements of $\bs C$ is Laplace. 
If the true model is not the LANOVA model and elements of $\bs C$ are drawn from a different heavy-tailed distribution, 
%
%$\widehat{\sigma}^2_c$ may not converge to the variance of elements of $\bs C$ and even if $\widehat{\sigma}^2_c$ does converge to the variance of $\bs C$,
%our posterior mode estimate of $\bs M$ may not correspond to the mode under the ``true'' prior distribution. 
it is natural to ask how our estimates $\widehat{\sigma}^2_c$, $\widehat{\sigma}^2_z$, and $\widehat{\bs M}$ perform.
As $\bs M$ is a function of $\mu$, $\bs a$, $\bs b$ and $\bs C$, the performance of $\widehat{\bs M}$ also reflects the performance of $\widehat{\mu}$, $\widehat{\bs a}$, $\widehat{\bs b}$ and $\widehat{\bs C}$ indirectly.
We find that the excess kurtosis $\kappa$ of the ``true'' distribution of elements of $\bs C$ determines our ability to estimate $\sigma^2_c$ separately from $\sigma^2_z$. 

\begin{proposition}\label{prop:misspec}
Under the model $\bs Y = \mu \bs 1_n \bs 1_p' + \bs a \bs 1_p' + \bs 1_n \bs b' + \bs C + \bs Z$, where elements of $\bs C$ are independent, identically distributed draws from a mean zero, symmetric distribution with variance $\sigma^2_c$, excess kurtosis $\kappa$ and finite eighth moment and elements of $\bs Z$ are normally distributed with mean zero and variance $\sigma^2_z$, $\widehat{\sigma}^2_c \stackrel{p}{\rightarrow}\sqrt{\kappa/3}\sigma^2_c$ and $\widehat{\sigma}^2_z\stackrel{p}{\rightarrow} \sigma^2_z + \left(1 - \sqrt{\kappa/3}\right)\sigma^2_c$ as $n \rightarrow \infty$ with $p$ fixed, $p \rightarrow \infty$ with $n$ fixed, or $n$ and $p\rightarrow \infty$.
\end{proposition}

Proposition~\ref{prop:misspec} indicates that we underestimate $\sigma^2_c$ when elements of $\bs C$ are lighter-than-Laplace tailed and we overestimate $\sigma^2_c$ when elements of $\bs C$ are heavier-than-Laplace tailed. To see how this affects estimation of $\bs M$, we consider exponential power and Bernoulli-normal distributed elements of $\bs C$. Exponential power distributed $c_{ij}$ have density $p(c_{ij} | \sigma^2_c, q_c) = (q_c/(2\sigma_c))\sqrt{\Gamma(3/q_c)/\Gamma(1/q_c)^3} \text{exp}\{- (\Gamma(3/q_c)/\Gamma(1/q_c))^{q_c/2} |c_{ij}/\sigma_c|^{q_c}\}$ that is parameterized in terms of the variance $\sigma^2_c$ and a shape parameter $q_c$, and Bernoulli-normal distributed $c_{ij}$ are exactly equal to zero with probability $1 - \pi_c$ and normally distributed with mean zero and $\tau^2_c$ otherwise. Both distributions can be lighter- or heavier-than-Laplace tailed. The excess kurtosis of exponential power and Bernoulli-normal $c_{ij}$ is $\Gamma\left(5/q_c\right)\Gamma\left(1/q_c\right)/\Gamma\left(3/q_c\right)^2 - 3$ and $3\left(1 - \pi_c\right)/\pi_c$, respectively. As a result, elements of $\bs C$ will be heavier-than-Laplace tailed when $q_c < 1$ or $\pi_c < 0.5$ and lighter-than-Laplace tailed when $q_c > 1$ or $\pi_c > 0.5$.
Note that when $q_c = 1$ the exponential power distribution corresponds to the Laplace distribution and the LANOVA model is correct, and when $\pi_c = 0.5$ the spike-and-slab distributed $\bs C$ have the same excess kurtosis as Laplace distributed $\bs C$ and the variance estimators $\widehat{\sigma}^2_c$ and $\widehat{\sigma}^2_z$ will be consistent. %% 

We compare the risk of the LANOVA estimate $\widehat{\bs M}$ to the risk of the maximum likelihood estimate $\widehat{\bs M}_{MLE}$, the risk of the strictly additive estimate $\widehat{\bs M}_{ADD}$, the risk of additive-plus-low-rank estimates $\widehat{\bs M}_{LOW,1}$ and $\widehat{\bs M}_{LOW,5}$ which assume rank-one and rank-five $\bs C$ respectively, the risk of the soft- and hard-thresholding IPOD estimates of \cite{She2011} $\widehat{\bs M}_{IPOD,S}$ and $\widehat{\bs M}_{IPOD,H}$, and the risk of approximately minimax estimates $\widehat{\bs M}_{MINI,U}$ and $\widehat{\bs M}_{MINI,S}$ obtained using the universal threshold and Stein's unbiased risk estimate (SURE) described by \cite{Donoho1994}. 
Additive-plus-low-rank estimates are computed according to \cite{Johnson1972b}.
Approximately minimax estimates are computed according to $\widehat{\bs M}_{MINI,U} = \widehat{\bs M}_{ADD} + \widehat{\bs C}_{MINI,U}$ and $\widehat{\bs M}_{MINI,S} = \widehat{\bs M}_{ADD} + \widehat{\bs C}_{MINI,S}$, where $\widehat{\bs C}_{MINI,S}$ and $\widehat{\bs C}_{MINI,S}$ are obtained by applying the universal threshold or SURE soft thresholding methods given by \cite{Donoho1994} to elements of $\bs Y - \widehat{\bs M}_{ADD}$, as if elements of $\bs Y - \widehat{\bs M}_{ADD}$ were independent and identically distributed.

%In general, the risk of $\widehat{\bs M}$ is not available in closed form. 
We compute Monte Carlo estimates of the relative risks 
for $n = p = 25$, $\mu = 0$, $\bs a = \bs 0$, $\bs b = \bs 0$ and $\sigma^2_z = 1$.
For exponential power distributed $\bs C$, we vary $\sigma^2_c = \left\{1/2,1,2 \right\}$ and $q_c = \left\{0.1,\dots,1.9 \right\}$. 
For Bernoulli-normal distributed $\bs C$, we vary $\tau^2_c = \left\{1/2,1,2 \right\}$ and $\pi_c = \left\{0,0.1,\dots,0.9,1 \right\}$. For each $(\sigma^2_c, q_c)$ and $(\tau^2_c, \pi_c)$, the Monte Carlo estimate is based on 500 simulated $\bs Y$.

\begin{figure}[h]
\centering
\includegraphics[scale=0.85]{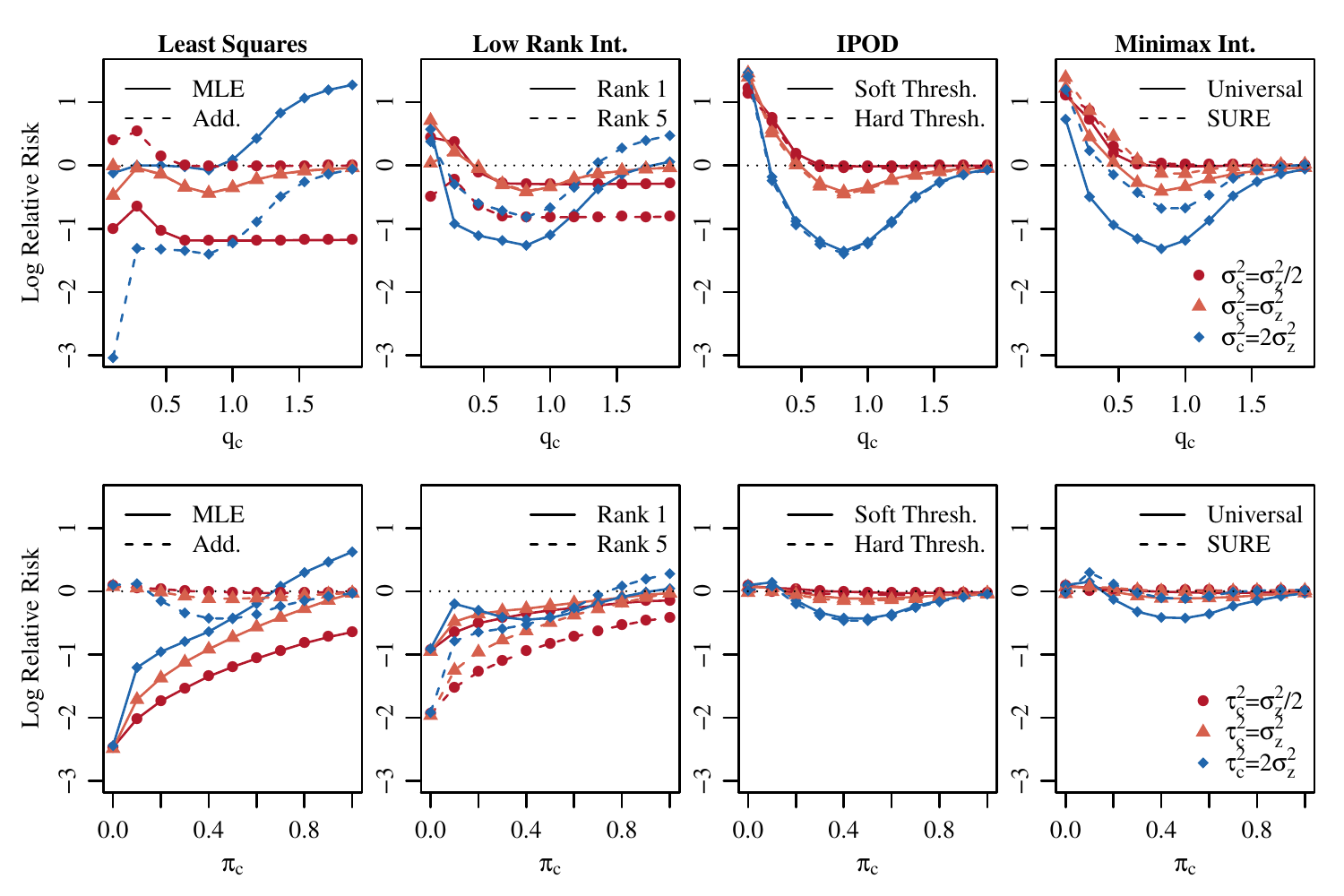}
\caption{Monte Carlo approximations of the relative risks of the LANOVA estimate $\widehat{\bs M}$ versus the MLE $\widehat{\bs M}_{MLE}$, the strictly additive estimate of $\widehat{\bs M}_{ADD}$, additive-plus-low-rank estimates $\widehat{\bs M}_{LOW,1}$ and $\widehat{\bs M}_{LOW,5}$ based on rank-1 and rank-5 estimates of $\bs C$, soft- and hard-thresholding IPOD estimates $\widehat{\bs M}_{IPOD,S}$ and $\widehat{\bs M}_{IPOD,H}$, and universal threshold and SURE based approximately minimax estimates $\widehat{\bs M}_{MINI,U}$ and $\widehat{\bs M}_{MINI,S}$.} \label{fig:misspec}
\end{figure}

The results shown in Figure~\ref{fig:misspec} indicate generally favorable performance of the LANOVA estimate $\widehat{\bs M}$. 
The top four plots show log relative risk estimates when elements of $\bs C$ are exponential power distributed, whereas the bottom four plots show log relative risk estimates when elements of $\bs C$ are Bernoulli-normal distributed.
 As expected, the LANOVA estimate performs as well as or better than all alternative estimators when $q_c = 1$ and the LANOVA model is true.
Interestingly, the LANOVA estimate also performs as well as or better than all alternative estimators when $\pi_c = 0.5$, even though the LANOVA model is not true. 
This suggests that the LANOVA estimate will tend to perform well relative to alternative estimators when the excess kurtosis of elements of $\bs C$ is similar to the excess kurtosis of a Laplace distribution.
This is consistent with Proposition~\ref{prop:misspec}, which states that the asymptotic bias of the variance estimators $\widehat{\sigma}^2_c$ and $\widehat{\sigma}^2_z$ will depend on the excess kurtosis of the true distribution of elements of $\bs C$.
 
In the first column of Figure~\ref{fig:misspec}, we see that the LANOVA estimate $\widehat{\bs M}$ tends to outperform the strictly non-additive estimate $\widehat{\bs M}_{MLE}$ when $q_c < 1$ or $\pi_c < 0.5$, especially when the variance of the interactions $\sigma^2_c$ or $\tau^2_c$ is small relative to the variance of the noise $\sigma^2_z$. Intuitively, this makes sense. When $q_c$ is small we expect that many elements of $\bs C$ will be very close to zero, and when $\pi_c$ is small, many elements of $\bs C$ will be exactly equal to zero. Furthermore, when $\sigma^2_c$ or $\tau^2_c$ are small we expect even the largest interactions to be small relative to the noise. Accordingly, this suggests that $\widehat{\bs M}$ tends to outperform the strictly non-additive estimate $\widehat{\bs M}_{MLE}$ when $\bs M$ is more nearly additive.
Analogously,  the LANOVA estimate $\widehat{\bs M}$ tends to outperform the strictly additive estimate $\widehat{\bs M}_{ADD}$ when $q_c > 1$ or $\pi_c > 0.5$, especially when the variance of the interactions $\sigma^2_c$ or $\tau^2_c$ is large relative to the variance of the noise $\sigma^2_z$. Again, this makes sense because we expect that fewer elements of $\bs C$ will be nearly or exactly equal to zero when $q_c > 1$ or $\pi_c > 0.5$, and more elements of $\bs M$ will be strictly non-additive.

In the second column of Figure~\ref{fig:misspec}, we see that the relative performance of the LANOVA estimate $\widehat{\bs M}$ relative to the additive-plus-low-rank estimates $\widehat{\bs M}_{LOW,1}$ and $\widehat{\bs M}_{LOW,5}$  depends on the distribution of elements of $\bs C$. 
When elements of $\bs C$ are exponential power distributed, the LANOVA estimate $\widehat{\bs M}$ outperforms the additive-plus-low-rank estimates $\widehat{\bs M}_{LOW,1}$ and $\widehat{\bs M}_{LOW,5}$  as long as $q_c$ is neither to small nor too large, especially when the variance of the interactions $\sigma^2_c$ is large relative to the variance of the noise $\sigma^2_z$. 
When elements of $\bs C$ are Bernoulli-normal distributed, the LANOVA estimate $\widehat{\bs M}$ almost always outperforms both additive-plus-low-rank estimates $\widehat{\bs M}_{LOW,1}$ and $\widehat{\bs M}_{LOW,5}$. This makes sense, as low rank approximations of sparse matrices tend to perform poorly.

In the third column, we see that the LANOVA estimate $\widehat{\bs M}$ tends to outperform the the soft- and hard-thresholding IPOD estimates $\widehat{\bs M}_{IPOD, S}$ and $\widehat{\bs M}_{IPOD, H}$ for values of $q_c > 0.5$ and all values of $\pi_c$ and $\tau^2_c$. 
%The LANOVA estimate $\widehat{\bs M}$ also tends to outperform the soft-thresholding IPOD estimate as long as $\pi_c$ is not very close to zero, especially when the variance of the nonzero interactions $\tau^2_c$ is large relative to the variance of the noise $\sigma^2_z$. 
The soft-thresholding IPOD estimate $\widehat{\bs M}_{IPOD, S}$ performs almost identically to the strictly additive estimate $\widehat{\bs M}_{ADD}$ which suggests that it tends to overpenalize elements of $\bs C$. Surprisingly, $\widehat{\bs M}_{IPOD,S}$ tends to slightly outperform $\widehat{\bs M}_{IPOD,H}$. As discussed in \cite{She2011}, outlier detection based on convex penalties, which are used to compute $\widehat{\bs M}$ and $\widehat{\bs M}_{IPOD,S}$, tends to perform worse than outlier detection based on nonconvex penalties, which are used to compute $\widehat{\bs M}_{IPOD,H}$. However the relatively better performance of $\widehat{\bs M}$ and $\widehat{\bs M}_{IPOD,S}$ relative to $\widehat{\bs M}_{IPOD, H}$ can be attributed to the fact that outlier detection based on convex penalties performs well when  all observations have equal leverage, as is the case in this setting \citep{Rousseeuw1987}.
The LANOVA estimate $\hat{\bs M}$ is also much faster to compute than both IPOD estimates; the soft- and hard-thresholding IPOD estimates take over $1,000$ times longer than the LANOVA estimate to compute on average across all simulations.

In the last column, we see that the LANOVA estimate $\widehat{\bs M}$ tends to outperform the approximately minimax estimates $\widehat{\bs M}_{MINI,U}$  and $\widehat{\bs M}_{MINI,S}$ as long as elements of $\bs C$ are not extremely heavy tailed or extremely sparse, especially when the variance of the interactions $\sigma^2_c$ is large relative to the variance of the noise $\sigma^2_z$.
Additionally, the improvements offered by the LANOVA estimate $\widehat{\bs M}$ over the approximately minimax estimates $\widehat{\bs M}_{MINI,U}$  and $\widehat{\bs M}_{MINI,S}$ are greater when $\bs C$ is exponential power distributed versus Bernoulli-normal distributed, which suggests that LANOVA penalization may be particularly useful when we expect that some elements of $\bs C$ are nearly but not exactly equal to zero.
%The approximately minimax estimates obtained by using the universal threshold $\widehat{\bs M}_{MINI,U}$ perform very similarly to the corresponding soft- and hard-thresholding IPOD estimates.
%The SURE based approximately minimax estimate $\widehat{\bs M}_{MINI,S}$ improves on the universal threshold. 

Altogether, these results also suggest that the LANOVA estimate $\widehat{\bs M}$ can offer better or comparable performance relative to alternatives as long as the tail behavior of elements of $\bs C$ is not too different from the tail behavior of a Laplace distribution.
Relatedly, the results also also suggest that we might be able to construct an improved estimate of $\bs M$ based on LANOVA penalization if prior information on the tail behavior of elements of $\bs C$ is available. The results displayed in Figure~\ref{fig:misspec} suggest that the biased estimation of $\widehat{\sigma}^2_c$ and $\widehat{\sigma}^2_z$ when elements of $\bs C$ have lighter- or heavier-than Laplace tailed described in Proposition~\ref{prop:misspec} leads to poorer LANOVA estimates of $\bs M$. Fortunately, this suggests that estimation of $\bs M$ could be improved if less biased estimates of $\widehat{\sigma}^2_c$ and $\widehat{\sigma}^2_z$ could be obtained. Proposition~\ref{prop:misspec} suggests a correction of multiplying $\widehat{\sigma}^2_c$ by $\sqrt{3/\kappa}$ and subtracting $(1 - \sqrt{\kappa/3})\sqrt{3/\kappa}\widehat{\sigma}^2_c$ from $\widehat{\sigma}^2_z$. % , could result in improved performance of the LANOVA estimate
However because excess kurtosis $\kappa$ is not a readily interpretable quantity, specifying a more appropriate value of $\kappa$ \emph{a priori} may be difficult. However if a Bernoulli-normal distribution for elements of $\bs C$ is plausible, a more appropriate value of $\pi_c$ can be used to specify a more appropriate value of $\kappa$. 
If the new value of $\pi_c$ is close to the ``true'' proportion of nonzero elements of $\bs C$, this may improve estimation of $\sigma^2_c$ and $\sigma^2_z$ and accordingly, $\bs M$.

\section{Extensions}\label{sec:ext}

%In this section, we describe two extensions to LANOVA penalization. The first extension involves penalizing lower-order parameters, $\bs a$ and $\bs b$, in addition to the higher order mean parameters, $\bs C$. The second involves tensor-variate $\bs Y$.

\subsection{Penalizing Lower-Order Parameters}\label{subsec:extrapen}

When $\bs Y$ has many rows or columns, it may be reasonable to believe that many elements of $\bs a$ or $\bs b$ are exactly zero.
A natural extension of Equation~\eqref{eq:obj} is given by
\begin{align}\label{eq:objlowpen}
\text{min}_{\mu,\bs a,\bs b, \bs C}\frac{1}{2\sigma^2_z}\left|\left|\text{vec}\left(\bs Y -\bs M\right) \right|\right|^2_2 + \lambda_a \left|\left|\bs a\right|\right|_1 + \lambda_b \left|\left|\bs b\right|\right|_1 + \lambda_c \left|\left| \text{vec}\left(\bs C\right)\right|\right|_1,
\end{align}
where we still have $\bs M = \bs 1_n \bs 1_p' \mu + \bs a \bs 1_p' + \bs 1_n \bs b' + \bs C$.
%We note that using different penalties for each group of parameters is somewhat similar to applying a group lasso to the groups of parameters, in that for sufficiently large values of $\lambda_a$, $\lambda_b$ or $\lambda_c$, an entire group of unknown mean parameters may be set to exactly zero.
Again, using the posterior mode interpretation of Equation~\eqref{eq:objlowpen}, we can estimate $\sigma^2_a$ and $\sigma^2_b$ from the observed data, $\bs Y$:
\begin{align*}
	\widehat{\sigma}^2_a &=\frac{1}{n-1}\sum_{i = 1}^n \check{a}^2_i - \frac{n}{\left(n -1\right)\left(p - 1\right)}\overline{r}^{\left(2\right)}\text{, \quad}\widehat{\sigma}^2_b = \frac{1}{p-1}\sum_{j = 1}^p \check{b}^2_j - \frac{p}{\left(n -1\right)\left(p - 1\right)}\overline{r}^{\left(2\right)}.
\end{align*}
where $\check{\bs a} = \bs H_n \bs Y \bs 1_p/p$ and $\check{\bs b} = \bs H_p \bs Y' \bs 1_n/n$ are OLS estimates for $\bs a$ and $\bs b$. The estimators $\widehat{\lambda}_a = \sqrt{2/\widehat{\sigma}^2_a}$ and $\widehat{\lambda}_b = \sqrt{2/{\widehat{\sigma}^2_b}}$ can be shown to be consistent for $\lambda_a$ and $\lambda_b$ as $n\rightarrow \infty$ and $p\rightarrow \infty$, respectively.
Because $\widehat{\lambda}_c$ and $\widehat{\sigma}^2_z$ do not depend on of $\bs a$ and $\bs b$, our estimators for $\lambda_c$ and $\sigma^2_z$ are unchanged.
%Given estimates of $\lambda_a$, $\lambda_b$, $\lambda_c$ and $\sigma^2_z$, Equation~\eqref{eq:objlowpen} is still a standard Lasso regression problem. 
A block coordinate descent algorithm for solving Equation~\eqref{eq:objlowpen} is given in the web appendix.
With respect to interpretation, population average row and column main effects can be obtained via ANOVA decomposition of $\widehat{\bs M}$.

\subsection{Tensor Data}

LANOVA penalization can be extended to a $p_1\times p_2 \times \dots \times p_K$ $K$-mode tensor $\bs Y$. We consider:
\begin{align}\label{eq:tenmod}
\text{vec}\left(\bs Y\right) &=\bs W \bs \beta + \text{vec}\left(\bs C\right)+ \text{vec}\left(\bs Z\right), \\ \nonumber
\bs C &= \left\{c_{i_1\dots i_K}\right\} \sim \text{i.i.d. Laplace}\left(0,\lambda_c^{-1}\right)\text{, \quad} \bs Z = \left\{z_{i_1 \dots i_K}\right\} \sim \text{i.i.d. N}\left(0,\sigma^2_z\right),
\end{align}
where $\text{vec}\left(\bs Y\right)$ is the $\prod_{k= 1}^K p_k \times 1$ vectorization of the $K$-mode tensor $\bs Y$ with ``lower'' indices moving ``faster'' and $\bs W$ and $\bs \beta$ are the design matrix and unknown mean parameters corresponding to a $K$-way ANOVA decomposition treating the $K$ modes of $\bs Y$ as factors. 
The matrix $\bs W = \left[\bs W_1, \dots , \bs W_{2^K-1}\right]$ is obtained by concatenating the $2^{K}-1$ unique matrices of the form $\bs W_l = \left(\bs W_{l,1} \otimes \dots \otimes \bs W_{l,K} \right)$,
where each $\bs W_{l,k}$ is equal to either $\bs I_{p_k}$ or $\bs 1_{p_k}$, excluding the identity matrix, $\bs I_{p_K}\otimes \dots \otimes \bs I_{p_1}$.
As in the matrix case, approaches that assume a low rank $\bs C$ are common \citep{VanEeuwijk1998, Gerard2015}. We penalize elements of the highest order mean term $\bs C$ for which no replicates are observed.
In the three-way tensor case, the first part of Equation~\eqref{eq:tenmod} refers to the following decomposition:
\begin{align}\label{eq:tensor}
y_{ijk} = \mu + a_i + b_j + d_k + e_{ij} + f_{ik} + g_{jk} + c_{ijk} + z_{ijk}.
\end{align}

%As in the matrix case, we suppose for a moment that no prior distribution were assumed for elements of $\bs C$.
Estimates of $\sigma^2_z$ and $\lambda_c$ are constructed from $\text{vec}\left(\bs R\right) = \left(\bs H_K \otimes \dots \otimes \bs H_1 \right)\text{vec}\left(\bs Y\right)$, where $\bs H_k = \bs I_{p_k} - \bs 1_{p_k} \bs 1_{p_k}/p_k$ is the $p_k \times p_k$ centering matrix and `$\otimes$' is the Kronecker product. 
As in the matrix case, $\text{vec}\left(\bs R\right)$ is independent of the lower-order unknown mean parameters $\bs \beta$, i.e. $\text{vec}\left(\bs R\right) = \left(\bs H_K \otimes \dots \otimes \bs H_1 \right)\text{vec}\left(\bs C + \bs Z\right)$. 
Our estimates of $\sigma^2_z$ and $\lambda_c$ are still functions of the second and fourth sample moments of $\bs R$: $\overline{r}^{\left(2\right)} = \frac{1}{p}\sum_{i = 1}^{p} r^2_{i}$ and $\overline{r}^{\left(4\right)} = \frac{1}{p}\sum_{i = 1}^{p} r^4_{i}$, where $p = \prod_{k = 1}^K p_k$.
We extend our empirical Bayes estimators as follows:
\begin{align}
\widehat{\sigma}^4_c =& \left\{\prod_{k = 1}^K\frac{p_k^3}{\left(p_k - 1\right)\left(p_k^2 - 3p_k + 3\right)}\right\}\left\{\overline{r}^{\left(4\right)}/3 - \left(\overline{r}^{\left(2\right)}\right)^2 \right\}, \\ \nonumber 
\widehat{\sigma}^2_c =& \sqrt{\widehat{\sigma}^4_c}\text{,\quad} \widehat{\sigma}^2_z = \left(\prod_{k = 1}^K\frac{p_k}{p_k - 1}\right)\overline{r}^{\left(2\right)} - \widehat{\sigma}^2_c,
\end{align}
where $\widehat{\lambda}_c = \sqrt{2/\widehat{\sigma}^2_c}$.  As in the matrix case, we can compute the bias of $\widehat{\sigma}^4_c$.
\begin{proposition}\label{prop:biastens}
Under the model given by Equation~\eqref{eq:tenmod}, 
\begin{align*}
\mathbb{E}\left[\widehat{\sigma}^4_c\right] -\sigma^4_c =& -\left\{\prod_{k = 1}^K\frac{p_k^3}{\left(p_k - 1\right)\left(p^2_k - 3p_k + 3\right)}\right\}\left[\left\{3\prod_{k = 1}^K\frac{\left(p_k - 1\right)^2}{p_k^3}\right\}\sigma^4_c +\right.\\
&\left. \left(2\prod_{k= 1}^K\frac{p_k - 1}{p_k^2}\right)\left(\sigma^2_c + \sigma^2_z\right)^2 \right].
\end{align*}
\end{proposition}
Interpretation of this result is analogous to the matrix case. We tend to prefer the simpler model with $\text{vec}\left(\bs C\right) = \bs 0$ over a more complicated model with nonzero elements of $\text{vec}\left(\bs C\right)$ when few data are available or when the data is very noisy.
Additionally,  $\mathbb{E}\left[\widehat{\sigma}^4_c\right] - \sigma^4_c=O\left(1/p\right)$, i.e. the bias of $\widehat{\sigma}^4_c$ diminishes as the number of levels of \emph{any} mode increases. We also assess the large-sample performance of our empirical Bayes estimators in the $K$-way tensor case.

\begin{proposition}\label{prop:consestsarr}
Under the model given by Equation~\eqref{eq:tenmod}, $\widehat{\sigma}^4_c\stackrel{p}{\rightarrow}\sigma^4_c$, $\widehat{\sigma}^2_c\stackrel{p}{\rightarrow}\sigma^2_c$, $\widehat{\lambda}_c\stackrel{p}{\rightarrow}\lambda_c$ and $\widehat{\sigma}^2_z\stackrel{p}{\rightarrow}\sigma^2_z$ as $p_{k'} \rightarrow \infty$ with $p_{k}$, $k\neq k'$, fixed or $p_1,\dots, p_K\rightarrow \infty$.
\end{proposition}

A block coordinate descent algorithm for estimating the unknown mean parameters is given in the web appendix.
Results for testing the appropriateness of assuming heavy-tailed $\bs C$ and robustness carry over to $K$-way tensors. 
$K$-way tensor analogues to Propositions~\ref{prop:test}-\ref{prop:misspec}, where we replace $np$ with $p$ and assume all $p_1,\dots,p_K\rightarrow \infty$, are shown to hold in the web appendix.
Lastly we can also extend LANOVA penalization for tensor data to penalize lower-order mean parameters.
Because tensor-variate $\bs Y$ include even more lower-order mean parameters, penalizing lower-order parameters is especially useful.
We give nuisance parameter estimators for penalizing lower-order parameters in the three-way case in the web appendix.

\section{Numerical Examples}\label{sec:numex}

\paragraph{Brain Tumor Data:}

We consider a $356\times 43$ matrix of gene expression measurements for 356 genes and 43 brain tumors. 
The 43 brain tumors include 24 glioblastomas and 19 oligodendrogal tumors, which include 5 astrocytomas, 8 oliodendrogliomas and 6 mixed oligoastrocytomas.
This data is contained in the \texttt{denoiseR} package for \texttt{R} \citep{Josse2016a}, and it has been used to identify genes associated with glioblastomas versus oligodendrogal tumors \citep{Bredel2005,deTayrac2009}. 
We focus on comparison to \cite{deTayrac2009}, who used a variation of principal components analysis of $\bs Y$ to identify differentially expressed genes and groups of tumors which is similar to using an additive-plus-low-rank estimate of $\bs M$. 
To ensure a straightforward comparison to the methods used in \cite{deTayrac2009}, we do not perform any additional preprocessing of the data, e.g. adjustments for possible low rank confounding factors that are often observed in gene expression data \citep{Leek2007, Gagnon-Bartsch2013}.

Unlike pairwise test-based methods which require prespecified tumor groupings, LANOVA penalization and additive-plus-low-rank estimates can be used to examine differential expression both within and across types of brain tumors. Differential expression \emph{within} types of brain tumors in particular is of recent interest \citep{Bleeker2012}.
%Accordingly, LANOVA penalization be used to identify genes that are differentially expressed across tumors as well as tumors with unique gene expression profiles by examining nonzero entries of $\widehat{\bs C}$.
%Differential expression is manifested as non-additivity of $\bs M$ to be non-additive and the corresponding elements of $\bs C$ to be nonzero. 
% Lots of variation within tumor type
%
 % Clarify
%We expect that the majority of the genes measured are not differentially expressed, i.e., many gene-by-tumor interaction effects are likely to be exactly zero.

%This data originally appeared in \cite{Bredel2005}. 

%Glioblastomas differ from oligodendroglial tumors in their invasiveness.
%Glioblastomas are the most invasive, while oligodendroglial tumors include several less invasive gliomas: astrocytomas, oliodendrogliomas, and mixed oligoastrocytomas.
%The original data measuring expression of 6,706 genes for 4 samples of healthy brain tissue, 31 glioblastomas and 14 oligodendroglial tumors was collected by  and used to identify genes that are differentially expressed in healthy brain tissue relative to glioblastomas or oligodendroglial tumors.

\begin{figure}[h]
\centering
\includegraphics[scale=0.85]{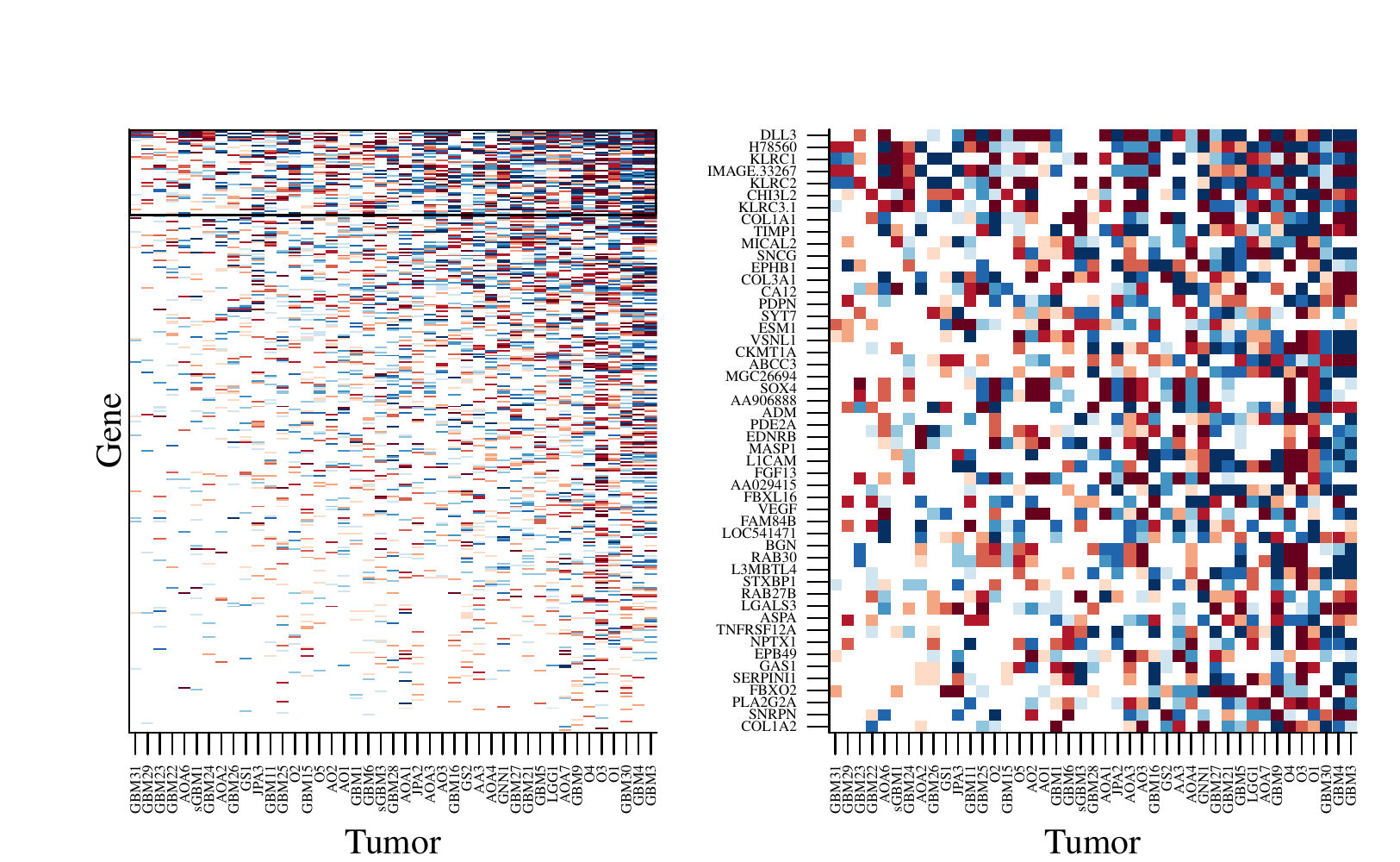}
\caption{Elements of the $356\times 43$ matrix $\widehat{\bs C}$. The first panel shows the entire matrix $\widehat{\bs C}$ with rows (genes) and columns (tumors) sorted in decreasing order of the row and column sparsity rates. The second panel zooms in on the rows of $\widehat{\bs C}$ (genes) marked in the first panel, which correspond to the fifty rows (genes) with the lowest sparsity rates. Colors correspond to positive (red) versus negative (blue) values of $\widehat{\bs C}$ and darkness corresponds to magnitude. 
%Darkness thresholds correspond to the $20\%, 40\%, 60\%, 80\%$ and $100\%$ quantiles of nonzero values of $|\widehat{\bs C}|$. 
}
\label{fig:prcomp} 
\end{figure}

We apply LANOVA penalization with penalized interaction effects and unpenalized main effects.
The test given by Proposition~\ref{prop:test} supports a non-additive estimate of $\bs M$; we obtain a test statistic of $18.45$ and reject the null hypothesis of normally distributed elementwise variability at level $\alpha = 0.05$ with  $p<10^{-5}$. % Exact value is $2.39\times e-76$
We estimate that $11,188$ elements of $\bs C$ (73\%) are exactly equal to zero, i.e. most genes are not differentially expressed. 
Figure~\ref{fig:prcomp} shows $\widehat{\bs C}$ and a subset containing fifty genes with the lowest gene-by-tumor sparsity rates. 
% This is in the caption, feels redundant:
%In both panels of Figure~\ref{fig:prcomp} tumors and genes are sorted in increasing order of the corresponding sparsity rates of the rows and columns of $\bs C$. 

%Now, we perform principal components analysis on $\widehat{\bs M}$ and examine the first two principal components, $\widetilde{\bs u}_1$ and $\widetilde{\bs u}_2$. Figure~\ref{fig:prcomp} shows that the $\widetilde{\bs u}_2$ separates glioblastomas and oligodendroglial tumors very well.
%This improves on the separation given by the second principal component of the raw data, $\bs Y $, shown in the second panel as well as the separation given by multiple factor analysis shown in \cite{deTayrac2009}.

The results of LANOVA penalization are consistent with those of \cite{deTayrac2009}. We observe that  49\% and 56\% of the elements of $\widehat{\bs C}$ involving the genes ASPA and PDPN are nonzero. 
Examination of $\widehat{\bs M}$  indicates overexpression of these genes among glioblastomas relative to oligodendrogal tumors, as observed in  \cite{deTayrac2009}. 
%
%%% Compare to previous research %%%
%%% Note: on previous research:
%%% - Bredel used SAM, a permutation based pairwise testing procedure with multiple testing adjustment (http://www.pnas.org/content/98/9/5116.full.pdf). All did normal brain vs. ??? (tumor, glioblastoma, other subtypes)
%%% - deTayrac used a kind of informed factor analysis which I think incorporates tumor types?
%
%% Bredel, Glioblastoma %%
%* BNIP2		Y	13
% HPRT1		Y	35
%* LYN		Y	51
% MSN		Y	53
% FCGR2B	Y	82		
%* CD44		Y	87
% ITGA5		Y	116
% HCLS-1	Y	123		HCLS1
%* CD47		Y	134
% IGFPB5	Y	140
% PLAU		Y	160
%* CYR61	Y	188
% TGFBI		Y	240
% FN1		Y	247
% PLAUR		Y	262
% IGFBP3	Y	290
% VEGF		Y	325
% IGFBP2	Y	353		As IMAGE.33267
%
% CDC2		N
% PTTG1		N
% IGF-I		N			I'm guessing this should be IGF-II?
%* CD151		N
%* CTGF		N
%
% de Tayrac 
%
%% None of these are in our data
% MGC39606
% CEAL1
% IGSF3
% GYPA
% HNRNPG.T
% BCAM
% EGLN2
% KIAA1543
% ZNF226
% EDG1
% ZNF329
% DCLRE1B
% SBP1
% ZNF233
% LILRA1
% APOC1
% ZNF160
% DUS3L
% ZNF419B
%
% MSN		Y	53
% RUNX1		Y	55
% PPP3CB	Y	78
% ARPP.19	Y	94
% HO8563	Y	114
% C9orf48	Y	117
% RTN3		Y	133
% CLIC1		Y	176
% WASF1		Y	194
% S100A11	Y	214
% UBA52		Y	215
% VAMP2		Y	231
% AA398420	Y	260
% RALY		Y	261
% X37864		Y	274
% PDXP		Y	277
% EMP3		Y	287
% AA281932	Y	297
% ASPA		Y	316
% PDPN		Y	342
LANOVA penalization yields additional results that are consistent with the wider literature.
The gene DLL3 has the highest rate of gene-by-tumor interactions at 74\% and tends to be underexpressed in glioblastomas. This is consistent with findings of overexpression of DLL3 in brain tumors with better prognoses \citep{Bleeker2012}. 
The KLRC genes KLRC1, KLRC2 and KLRC3.1 all have very high rates of gene-by-tumor interactions at 72\%, 70\% and 60\%.
\cite{Ducray2008} has found evidence for differential KLRC expression across glioma subtypes.
%Last, 63\% of gene-by-tumor interactions involving the gene CHI3L2 are nonzero. This is consistent with research that has found differential expression of CHI3L2 when comparing primary and secondary glioblastomas \citep{Tso2006}. 
LANOVA penalization also indicates that several brain tumors have unique gene expression profiles. Glioblastomas 3, 4 and 30 have rates of nonzero gene-by-tumor interactions exceeding 50\% and similar gene expression profiles. 
Specifically, we observe overexpression of FCGR2B and HMOX1 and underexpression of RTN3 for gliomastomas 3, 4 and 30. Overexpression of FCGR2B or HMOX1 is associated with poorer prognosis \citep{Zhang2016, Ghosh2016a}, and RTN3 is differentially expressed across subgroups of glioblastoma that differ with respect to prognosis \citep{Cosset2017}.
This suggests that glioblastomas 3, 4 and 30 may correspond to an especially aggressive subtype.

\paragraph{fMRI Data:}

Second, we consider a tensor of fMRI data which appeared in \cite{Mitchell2004}.
During each of 36 tasks, fMRI activations were measured at $55$ time points and $4,698$ locations (voxels).
Accordingly, the data can be represented as a $36\times 55\times 4,698$ three-way tensor.
Because the data is so high dimensional, many methods of analysis are prohibitively computationally burdensome. 
Accordingly, parcellation approaches that reduce the spatial resolution of fMRI data by grouping voxels into spatially contiguous groups are common, however, the choice of a specific parcellation can be difficult \citep{Thirion2014}.
Instead, we propose LANOVA penalization as an exploratory method to identify relevant dimensions of spatial variation that should be accounted for in a subsequent analysis. 

The test given by Proposition~\ref{prop:test} supports a non-additive estimate of $\bs M$; we obtain a test statistic of $298.87$ and reject the null hypothesis of normally distributed elementwise variability at level $\alpha = 0.05$ with  $p<10^{-5}$. %10^{-16}
Having found support for the use of a non-additive estimate of $\bs M$, we also penalize lower-order mean parameters as $\bs Y$ is high dimensional and sparsity of lower-order mean parameters could result in substantial dimension reduction and improved interpretability.
The LANOVA estimate has $1,751,179$ nonzero parameters, a small fraction of the $9,302,040$ parameters needed to represent the raw data, $\bs Y$ ($18.83\%$). %%

\begin{figure}[h]
\centering
\includegraphics[scale = 0.85]{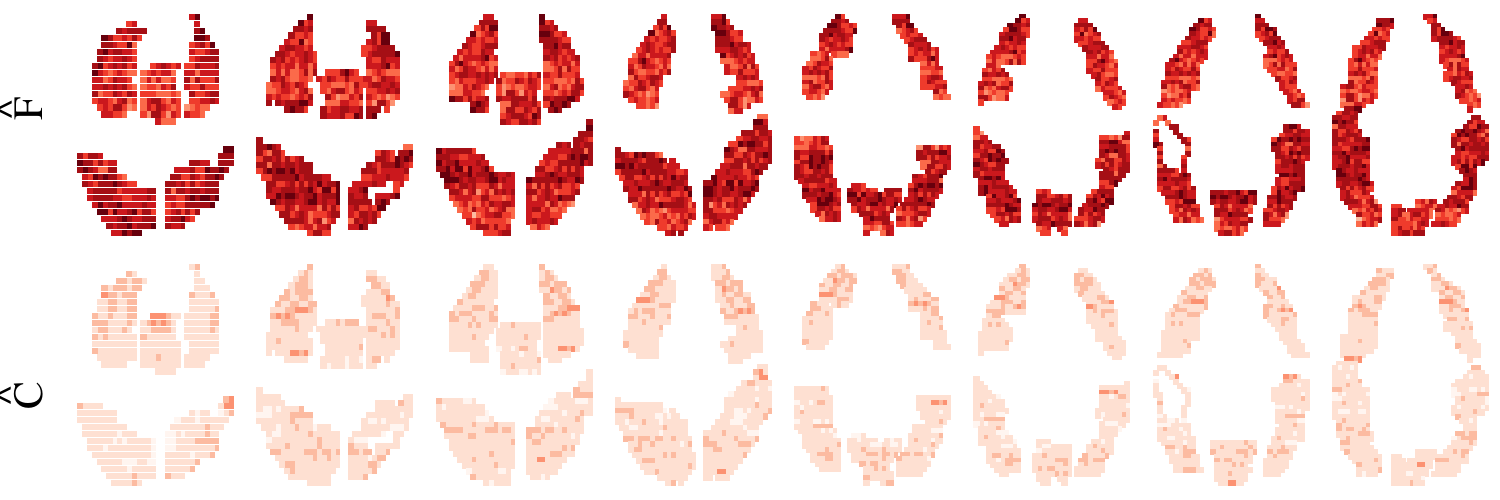}
\caption{Percent of nonzero entries of $\widehat{\bs F}$ and $\widehat{\bs C}$ by location, where $\widehat{\bs F}$ and $\widehat{\bs C}$ estimate $\bs F$ and $\bs C$ as defined in Equation~\eqref{eq:tensor}. Entries of $\bs F$ index task-by-location interaction terms and entries of $\bs C$ index task-by-time-by-location elementwise interaction terms. Darker colors indicate higher percentages.}
\label{fig:fmri} 
\end{figure}

Recalling that we are primarily interested in spatial variation, we examine estimated task-by-location interactions $\widehat{\bs F}$ and task-by-time-by-location elementwise interactions $\widehat{\bs C}$,  as defined in Equation~\eqref{eq:tensor}.
Figure~\ref{fig:fmri} shows the percent of nonzero entries $\widehat{\bs F}$ and $\widehat{\bs C}$ at each location.  
At each location, the proportion of nonzero entries of $\widehat{\bs F}$ is much larger than the proportion of nonzero entries of $\widehat{\bs C}$. This suggests that much of the spatial variation of activations by task can be attributed to an overall level change in activation over the duration of the task, as opposed to time-specific changes in activation. In a subsequent analysis, it may be reasonable to ignore task-by-time-by-location interactions.

%For both $\widehat{\bs F}$ and $\widehat{\bs C}$, we see spatial patterns in the percent of nonzero entries that offer some insight into the regions that are most responsive to the tasks performed in this data.
By examining the percent of nonzero entries of $\widehat{\bs F}$ by location, we can get a sense of which locations correspond to level changes in fMRI activity response by task.
There is evidence for an overall level change in response to at least some tasks for all locations; the minimum percent of nonzero entries of $\widehat{\bs F}$ per location is $33\%$. However, voxels in the the parietal region, the calcarine fissure and the right- and left-dorsolateral prefrontal cortex have particularly high proportions of nonzero entries of $\widehat{\bs F}$, suggesting that overall activation in these regions is particularly responsive to tasks.
By examining the percent of nonzero entries of $\widehat{\bs C}$ by location, we can get a sense of which locations correspond to time-specific differential activity by task over time. 
We see that nonzero entries of $\widehat{\bs C}$ are concentrated among voxels in the upper supplementary motor area, the calcarine fissure and the left- and right-temporal lobes.
In this way, we can use LANOVA estimates to identify subsets of relevant voxels that should be included in a subsequent analysis.

\paragraph{Fusarium Data:}

Last, we consider the problem of checking for nonzero three-way interactions in experimental data without replicates.
The data is a $20\times 7 \times 4$ three-way tensor containing severity of disease incidence ratings for 20 varieties of wheat infected with 7 strains of Fusarium head blight over 4 years, from 1990-1993 that appeared in \cite{VanEeuwijk1998}.
There is scientific reason to believe that several nonzero three-way variety-by-strain-by-year interactions are present. \cite{VanEeuwijk1998} examined these interactions using a rank-one tensor model for $\bs C$, i.e.\ $c_{ijk} = \alpha_{i} \gamma_{j} \delta_{k}$. However as noted in Section~\ref{sec:intro}, a low rank model may not be sufficient even if few nonzero interactions are present. 

As in \cite{VanEeuwijk1998}, we transform the severity ratings to the logit scale before estimating LANOVA parameters.
The test given by Proposition~\ref{prop:test} supports a non-additive estimate of $\bs M$; we obtain a test statistic of $3.99$ and reject the null hypothesis of normally distributed elementwise variability at level $\alpha = 0.05$ with $p=3.34\times10^{-5}$. %$3.340176e-05$
We obtain 87 nonzero entries of $\widehat{\bs C}$ ($16\%$). 

\begin{figure}[h]
\centering
\includegraphics[scale = 0.85]{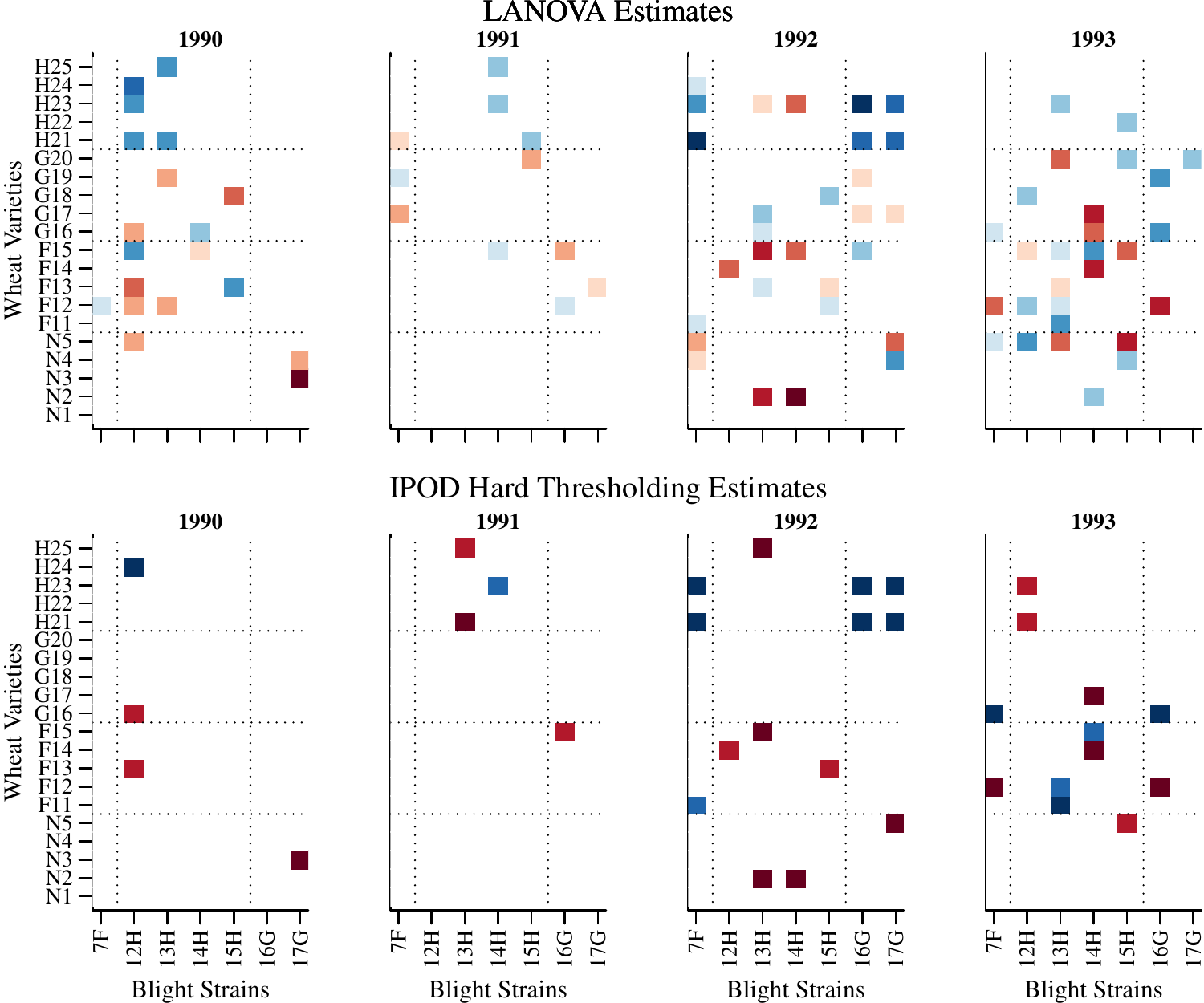}
\caption{Entries of $\widehat{\bs C}$. Red (blue) points indicate positive (negative) nonzero entries of $\widehat{\bs C}$ and darker colors correspond to larger magnitudes.}
\label{fig:fusarium} 
\end{figure}

Figure~\ref{fig:fusarium} shows nonzero elements of $\widehat{\bs C}$ as well as nonzero elements of $\widehat{\bs C}_{IPOD,H}$ obtained using the hard-thresholding IPOD method of \cite{She2011}, with dashed lines separating groups of wheat and blight by country of origin: Hungary (H), Germany (G), France (F) or the Netherlands (N).
Estimates of elements of $\widehat{\bs C}_{IPOD,S}$ obtained using the soft-thresholding IPOD method of \cite{She2011} are not pictured because they are all exactly equal to zero.
We can interpret nonzero elements of estimates of $\bs C$ as evidence for variety-by-year-by-strain interactions that cannot be expressed as additive in variety-by-year, year-by-strain and variety-by-strain effects.
Like \cite{VanEeuwijk1998}, both the LANOVA estimate $\widehat{C}$ and the hard-thresholding IPOD estimate $\widehat{\bs C}_{IPOD, H}$ include large three-way interactions in 1992, during which there was a disturbance in the storage of blight strains. Specifically, we observe interactions involving Dutch variety 2, the only variety with no infections at all in 1992, and interactions between Hungarian varieties 21 and 23 and foreign blight strains, which despite the storage disturbance were still able to cause infection in these two Hungarian varieties alone.
The soft-thresholding IPOD estimate $\widehat{\bs C}_{IPOD,S}$ fails to include these real interactions, suggesting that it overpenalized $\bs C$ just as we observed in simulations.
We do not have enough information about the data to assess whether or not the remaining interactions identified by $\hat{\bs C}$ and $\hat{\bs C}_{IPOD,H}$ are related to features of the study or known patterns in the behavior of certain varieties and strains, however they suggest further investigation of these varieties and strains may be warranted.

\section{Discussion}\label{sec:disc}

This paper demonstrates the use the common Lasso penalty and the corresponding Laplace prior distribution for estimating elementwise effects of scientific interest in the absence of replicates.
Our procedure, LANOVA penalization, can also be interpreted  as assessing evidence for nonadditivity.
We show that our nuisance parameter estimators are consistent and explore their behavior when assumptions are violated. We demonstrate that  the corresponding mean parameter estimates can perform favorably relative to strictly additive, strictly non-additive,  additive-plus-low-rank, IPOD, and approximately minimax estimates when elements of $\bs C$ are exponential power or Bernoulli-normal distributed, especially if the tail behavior of elements of $\bs C$ is similar to the tail behavior of a Laplace distribution and the variance of the interactions $\sigma^2_c$ is large relative to the variance of the noise $\sigma^2_z$.
We emphasize that LANOVA penalization is computationally simple. 
The nuisance parameter estimators are easy to compute for arbitrarily large matrices and estimates of $\bs M$ can be computed using fast a block coordinate descent algorithm that exploits the structure of the problem.
We also extend LANOVA penalization to penalize lower-order mean parameters and apply to tensors.
Finally, we show that LANOVA estimates can be used to examine gene-by-tumor interactions using microarray data, to perform exploratory analysis of spatial variation in activation response to tasks over time in high dimensional fMRI data and to assess evidence for ``real'' elementwise interaction effects at  in experimental data. To conclude, we discuss several limitations and extensions.

One limitation is that we assume heavy-tailed elementwise variation is of scientific interest and should be incorporated into the mean $\bs M$, whereas normally distributed elementwise variation is spurious noise $\bs Z$. If the noise is heavy-tailed, it may erroneously be incorporated into the estimate of $\bs C$. 
Similar issues arise with low rank models for $\bs C$, insofar as systematic correlated noise can erroneously be incorporated into the estimate of $\bs C$. 
Furthermore, if the elementwise variation of scientific interest includes low rank components, these low rank components may not be included in LANOVA estimates of the mean $\bs M$. This is of particular concern in the brain tumor data example but also of more general concern in applications involving gene expression data, because unobserved confounders with low rank structure may be present and need to be accounted for \citep{Leek2007, Gagnon-Bartsch2013}.
That said, any method that aims to separate elementwise variation into components that are of scientific interest and spurious noise requires strong assumptions that must be considered in the context of the problem. 

Another limitation is that the results of the simulation study assessing the performance of the LANOVA estimate $\widehat{\bs M}$ do not necessarily suggest favorable performance of the LANOVA estimate in all settings. First, we only consider exponential power and Bernoulli-normal elements of $\bs C$. Although we expect the LANOVA estimate to perform well when elements of $\bs C$ are heavy tailed, we do not expect the LANOVA estimate to perform well when $\bs C$ are light-tailed. Second, this scenario considers $\bs C$ with independent, identically distributed elements. In some settings, it may be reasonable to expect dependence across elements of $\bs C$ and additive-plus-low-rank estimates may perform better.

The methods presented in this paper could be extended in several ways.
Although we use our estimators of $\lambda_c$ and $\sigma^2_z$ in posterior mode estimation, the same estimators could also be used to simulate from the posterior distribution using a Gibbs sampler.
The output of a Gibbs sampler could be used to construct credible intervals for elements of $\bs M$, which would be one way of addressing uncertainty.
We may also want to account for additional uncertainty induced by estimating $\lambda_c$ and $\sigma^2_z$.
To address this, a fully Bayesian approach with prior distributions set for $\lambda_c$ and $\sigma^2_z$ could be taken at the cost of losing a sparse estimate of $\bs C$.
Our empirical Bayes nuisance parameter estimators could be used to set parameters of prior distributions for $\lambda_c$ or $\sigma^2_z$.
Last, LANOVA penalization for matrices is a specific case of the more general bilinear regression model, where we assume $\bs Y = \bs A \bs W + \bs B \bs X + \bs C + \bs Z$, given known $\bs W$ and $\bs X$. We chose to focus on a simpler case in this paper to facilitate the derivation of interpretable expressions for the bias of $\sigma^4_c$ as well as estimators that are consistent as either $n$ or $p \rightarrow \infty$ because researchers often encounter ``fat'' or ``skinny'' matrices in practice. However, the same logic could easily be extended to the more general bilinear regression context as well as the even more general multilinear context for tensor $\bs Y$.

Finally, our intuition combined with the results of Section~\ref{sec:laimp} suggests that the Laplace distributional assumptions used in this paper are likely to be violated in many settings.
The strength of the Laplace distributional assumption for elements of $\bs C$ can be justified by the need to make \emph{some} assumptions to estimate $\sigma^2_c$ and $\sigma^2_z$ when $\bs C$ and $\bs Z$ are always observed as a sum $\bs C + \bs Z$. Given that we need to estimate fourth order moments of $\bs C$ and $\bs Z$  just to separately estimate $\sigma^2_c$ and $\sigma^2_z$, we expect that we would need to estimate even higher order moments to assess the appropriateness of the Laplace distributional assumption for $\bs C$ which is notoriously difficult to do well in practice. However, if we observed replicate measurements, as is the case for lower-order mean parameters $\bs a$ and $\bs b$, we would not need to estimate fourth order moments to separately estimate $\sigma^2_a$ or $\sigma^2_b$ from $\sigma^2_z$. Instead, we could use fourth order moments to assess the appropriateness of the Laplace distributional assumptions for $\bs a$ or $\bs b$ and possibly improve distribution specification.
In recent work, we have considered this question in the general regression setting and have found that it can be possible to test the appropriateness of Laplace distributional assumptions and, when Laplace distributional assumptions are inappropriate, choose better ones \citep{Griffin2017c}. % **

\section*{Acknowledgements}
This work was partially supported by NSF grants DGE-1256082 and DMS-1505136.

\addcontentsline{toc}{section}{References}
\bibliography{LANOVA.bib}
\bibliographystyle{chicago}

\end{document}